\documentclass[12pt,draftclsnofoot,onecolumn]{IEEEtran}

\usepackage{amsfonts}
\usepackage{mathrsfs}
\usepackage[noblocks]{authblk}
\usepackage[compress]{cite}
\usepackage{bm}
\usepackage{amsmath,amssymb,epsfig,graphics,subfigure}
\usepackage{theorem}
\usepackage{array,color}
\usepackage{algorithm}
\usepackage{algorithmic}
\usepackage[OT1]{fontenc}

\newtheorem{conclusion}{Conclusion}

\allowdisplaybreaks
\theoremheaderfont{\normalfont\bfseries}

\usepackage{float}
\usepackage[dvipsnames]{xcolor}

\newcommand{\Hm}{\mathrm{H}}

\begin{document}

\title{\huge Hybrid Transceiver Optimization for Multi-Hop Communications}
 
 \author{Chengwen Xing,~\textsl{Member, IEEE}, Xin Zhao, Shuai Wang, Wei Xu,~\textsl{Senior Member, IEEE},
 	{Soon~Xin~Ng,~\textsl{Senior~Member,~IEEE}}, Sheng Chen,~\textsl{Fellow, IEEE}
 	\thanks{C. Xing, X. Zhao and S. Wang are with School of Information and
 		Electronics, Beijing Institute of Technology, Beijing 100081, China
 		(e-mail: chengwenxing@ieee.org; xinzhao.eecs@gmail.com; swang@bit.edu.cn).}%
 	\thanks{W.~Xu is with the National Mobile Communications Research
 		Laboratory, Southeast University, Nanjing, China (e-mail: wxu@seu.edu.cn).}%
 	\thanks{S.~X.~Ng and S.~Chen are with School of Electronics and Computer
 		Science, University of Southampton, U.K. S. Chen is also with King Abdulaziz
 		University, Jeddah, Saudi Arabia (e-mail: sxn@ecs.soton.ac.uk; sqc@ecs.soton.ac.uk).}%
 	\vspace*{-16mm}
 }
\maketitle
		
\begin{abstract}
 Multi-hop communication with the aid of large-scale antenna arrays will play a
 vital role in future emergence communication systems. In this paper, we investigate
 amplify-and-forward based and multiple-input multiple-output assisted multi-hop
 communication, in which all nodes employ hybrid transceivers. Moreover, channel
 errors are taken into account in our hybrid transceiver design. Based on the
 matrix-monotonic optimization framework, the optimal structures of the robust
 hybrid transceivers are derived. By utilizing these optimal structures, the
 optimizations of analog transceivers and digital transceivers can be separated
 without loss of optimality. This fact greatly simplifies the joint optimization
 of analog and digital transceivers. Since the optimization of
 analog transceivers under unit-modulus constraints is nonconvex, a projection
 type algorithm is proposed for analog transceiver optimization to overcome this
 difficulty. Based on the derived analog transceivers, the optimal digital
 transceivers can then be derived using matrix-monotonic optimization. Numerical
 results obtained demonstrate the performance advantages of the proposed hybrid
 transceiver designs over other existing solutions.
\end{abstract}
	
		
\section{Introductions}\label{sect:intro}
	
 Emergency communications are of critical importance in managing emergency scenarios,
 such as natural disasters, anti-terrorist wars, large-scale sport events
 \cite{emergencyMultiHop2016}. Multi-hop communication is an important enabling
 technology for emergency communications because it is less demanding on network
 infrastructures. For example, multi-hop communications can occur between multiple
 satellites or multiple unmanned aerial vehicles or other high-latitude platforms
 \cite{satelliteTVT,UAVRelay2016,YRongTWC2014MultiHop}. Moreover, multi-hop 
 communication is also a promising technology to overcome deep fadings over long
 distance for high frequency band communications \cite{MaXiaoLiTWC2014},
 such as millimeter wave communications or Terahertz communications
 \cite{XingTSP2013,XingTSP201502,XingTSP2016,PoorJSAC2015,MaXiaoLiTWC2014}.

 Generally, it is challenging to simultaneously guarantee high reliability and
 high spectrum efficiency of multi-hop communications \cite{multiHopSurvey2013}.
 Because of its high spatial diversity and multiplexing gains, the large-scale
 antenna array technology offers a promising candidate for this difficult task.
 It is worth highlighting that different from cellular communications, the
 physical-size constraints on emergence communication nodes are less stringent.
 As a result, it is practical to apply multiple-input multiple-output (MIMO)
 technology to overcome path loss and to improve spectral efficiency simultaneously.
 For MIMO multi-hop communications, various signal processing strategies at relays
 can be classified into two categories, i.e., regenerative operation and
 nonregenerative operation. For regenerative schemes, the signal received
 at each intermediate hop is decoded first, then a new transmission for the decoded information is performed to the next hop. For nonregenerative schemes,
 the received signal from the preceding hop is not decoded but directly forwarded
 to the next hop after multiplying it with a forward matrix. Nonregenerative
 schemes are characterized by their low complexity and high security \cite{XingTSP2016}.

 In order to meet the demands of data-hungry applications, the scale of MIMO has become increasingly larger, and the cost of antenna arrays in MIMO systems has boosted dramatically, correspondingly \cite{largeScaleMCOM2015}. In particular, the deployment of large-scale antenna arrays will inevitably be impeded by the significant cost and complexity of emergence communication nodes \cite{VariablePhaseTSP2005}. To get over the limitations due to the high cost and implementation complexity, hybrid analog/digital structures have been proposed, which have attracted lots of attention \cite{hybSurvey2016,hybTWC2016,HeathTWC2014}. Unlike the traditional full digital systems, in hybrid transceivers, part of signal processing work is delegated to radio-frequency (RF) devices, which could greatly reduce the cost of MIMO transceivers \cite{hybTWC2016,HeathTWC2014}. The subsequent challenges mainly come from the analog transceiver optimizations, because the unit-modulus constraints on each element of the analog transceiver matrices are nonconvex and difficult to solve using existing algorithms \cite{HeathTWC2014,hybSurvey2016,DongTSP2017}.
 
 The potential of hybrid transceivers in mmWave communications arouses a great passion in hybrid beamforming design. Existing literatures mainly concentrated on the topic of exploring and optimizing the hybrid beamforming strategies in varies communication systems \cite{hybSurvey2016,hybTWC2016,subArrayTCOM2018,subArrayLinJSAC2017}. The techniques in compress sensing were firstly introduced to deal with the point-to-point hybrid transceiver design in \cite{HeathTWC2014}. The authors in \cite{DongTSP2017,WeiYuHybJSTSP2016,JZhangJSTSP2016Alternating} improved the performance of the point-to-point hybrid communication systems with the sacrifice of higher computational complexity. Then, MSE criterion based hybrid design and selection based hybrid structure were investigated in \cite{JZhangMSETCOMM2019,EldarHybTSP2019,HybirdSelectionTSP2018}. The investigations were not only limited to the point-to-point linear transceiver optimizations \cite{heathLimit2015,robustLCOMM2016}. In fact, the nonlinear hybrid transceiver optimizations have drew more attentions recently. The nonlinear hybrid transceiver with vector perturbation for the point-to-point communication was studied in \cite{MaiTVT2019nonlinear}. A general nonlinear hybrid transceiver optimization was discussed in \cite{matrixMonoOpt2018}. Further, the hybrid transceiver design in multi-user and multi-cell communications became one of the major concerns in hybrid beamforming topic \cite{MYHongJSTSP2017Hybrid,hybMUJSAC2017,channelStatTWC2017,upLinkHybTVT2017,WXuTWC2018MU_MIMO,gong2019hybrid}. The work in \cite{MultiCellTWC2018} proposed an analytical framework for multi-cell hybrid communications, while only single stream mmWave communications were thoroughly investigated. Afterwards, the hybrid precoding optimization was naturally extended to the dual-hop relay communications \cite{hybridRelay2014,hybRelayTCOM2017,DaiLLRelayTSP2018,QJShiTWC2019Duplex}. In \cite{hybridRelay2014}, the authors also tried to use compress sensing based algorithm to handle the analog relaying beamformer design with limitation of mmWave channels. The work in \cite{hybRelayTCOM2017} considered the dual-hop relay system with two-hop relaying strategy, which can be applied to the massive MIMO channels. Other researchers investigated the full duplex two-hop relay communications based on the nonconvex optimization algorithms \cite{QJShiTWC2019Duplex}. However, to the best of the authors' knowledge, few has taken into account the hybrid transceiver optimization in relay communications, e.g., for emergency communication scenarios. Neither the multi-hop hybrid relay communication nor a general framework of hybrid relay communications has ever been reported.
 

 Different from these existing works, in this work, we investigate the hybrid transceiver
 designs for a multi-hop AF MIMO cooperative network \cite{XingTSP201501,matrixMonoOpt2018}.
 Furthermore, channel errors are also taken into account \cite{XingIET2013}. More
 specifically, we propose a comprehensive unified framework of robust hybrid transceiver
 optimizations for multi-hop cooperative communications. Our work is much more
 challenging than the existing works. The main contributions of this work are listed as
 follows, which differentiate our work from the existing works distinctly.
\begin{itemize}
\item We consider a general multi-hop AF MIMO relaying system, where multiple
 relays facilitate the communications between source and its destination. All
 nodes are equipped with multiple antennas and multiple data streams are simultaneously
 transmitted. In addition, both linear transceivers and nonlinear transceivers are
 investigated in our framework. The nonlinear transceivers investigated include
 Tomlinson-Harashima precoding (THP) at the source or decision feedback equalizer (DFE)
 at the destination \cite{bookTHP,XingJSAC2012,DFETSP2004}.
\item For the linear transceiver designs of the multi-hop AF MIMO relaying network, two
 general types of performance metrics are considered, namely, additively Schur-convex
 function and additively Schur-concave function of the diagonal elements of the data
 estimation matrix at the destination. Different fairness levels can be realized by
 using these two types of performance metrics.
\item For the nonlinear transceiver designs of the multi-hop AF MIMO relaying network,
 two general kinds of performance metrics are considered, namely, multiplicatively
 Schur-convex function and multiplicatively Schur-concave function of the diagonal
 elements of the data estimation matrix at the destination. Different fairness levels
 can be compromised by leveraging these two kinds of performance metrics.
\item In our work, correlated channel errors in each hop are taken into account.
 The correlated channel errors make the hybrid transceiver optimization for AF
 MIMO relaying networks particularly challenging, and to the best of our knowledge,
 this robust hybrid transceiver optimization has not be addressed in existing
 literature.
\item At source and destination, the hybrid transceiver consists of two parts, i.e.,
 analog and digital precoders as well as analog and digital receivers, respectively.
 At each relay, the hybrid transceiver consists of three components, i.e., analog
 receive part, digital forward part and analog transmit part. Based on the
 matrix-monotonic framework \cite{XingTSP201501}, the optimal structures of these
 components are derived. By exploiting these optimal structures, the robust hybrid
 transceiver for multi-hop communications is optimized efficiently. Our results
 can be applied to many frequency bands, including RF, millimeter wave and Terahertz
 bands.
\end{itemize}
	
 Throughout our discussions, bold-faced lower-case and upper-case letters denote
 vectors and matrices, respectively. The Hermitian square root of a positive
 semi-definite matrix $\bm{M}$ is denoted by $\bm{M}^{\frac{1}{2}}$. The expectation
 operator is denoted by $\mathbb{E}\{\cdot\}$ and ${\rm Tr}(\cdot )$ is the matrix
 trace operator. While $(\cdot )^{\rm T}$, $(\cdot )^{*}$, $(\cdot )^{\rm H}$ and
 $(\cdot )^{-1}$ denote matrix transpose, conjugate, Hermitian transpose and inverse
 operators, respectively. The diagonal matrix with the  diagonal elements
 $\lambda_1, \cdots ,\lambda_N$ is denoted as ${\rm diag}\{\lambda_1, \cdots ,
 \lambda_N\}\! =\! {\rm diag}\big\{[\lambda_1 \cdots \lambda_N]^{\rm T}\big\}$, and
 $\bm{I}$ denotes the identity matrix with appropriate dimension, while $\bm{d}[\bm{M}]$
 is the vector whose elements are the diagonal elements of matrix $\bm{M}$, and
 $\bm{d}^2[\bm{M}]\! =\! \bm{d}\big[{\rm diag}\{\bm{d}[\bm{M}]\}{\rm diag}^*\{\bm{d}[\bm{M}]\}\big]$.
 The real part operation is denoted by $\Re\{\cdot\}$, and the angle of scalar
 $a$ is denoted as $\measuredangle a$. The symbol $\mathcal{P}_{\mathcal{F}}\{\cdot\}$
 denotes the angle projection operation, i.e., $\mathcal{P}_{\mathcal{F}}\{a\}\! =\!
 e^{\textsf{j} \cdot\measuredangle a}$, where $\textsf{j}\! =\! \sqrt{-1}$, and $\|\cdot\|_F$
 is the matrix Frobenius norm. $\bm{\Lambda}\searrow$ represents a rectangular
 or square diagonal matrix whose diagonal elements are arranged in decreasing order,
 while $\bm{\lambda}\{\bm{M}\}\! =\! [\lambda_1(\bm{M}) ~ \lambda_2(\bm{M})\cdots
\lambda_N(\bm{M})]^{\rm{T}}$, where $\lambda_n(\bm{M})$ is the $n$th largest eigenvalue
 of the $N\times N$ matrix $\bm{M}$. Furthermore, $(a)^\dag\! =\! \max\{0,a\}$.
 `Independently and identically distributed' and `with respect to' are abbreviated as
 `i.i.d.' and `w.r.t.', respectively.
	
\section{System Model and Problem Formulation}\label{S2}
	
 Consider a general multi-hop ($K$-hop) AF MIMO relaying network in which multiple ($K\!
 -\! 1$) relay nodes (nodes 2 to $ K $) help a source node (node 1) to communicate with a destination node (denoted as node $ K $). At each relay, the
 received signal vector is not decoded but is directly forwarded to the next node after
 multiplying it with a forward matrix. All the nodes are equipped with multiple antennas
 and multiple data streams are simultaneously transmitted. Define the number of transmit
 and receive antennas at the $ k $th node as $N_{t,k}$ and $N_{r,k}$, the number of
 RF chains in the structure as $N_{\rm RF}$, and the number of data streams as $N$. Let
 the transmitted signal vector from the source be $\bm{x}_0\! \in\! \mathbb{C}^N$ with
 $\mathbb{E}\big\{\bm{x}_0\bm{x}_0^{\rm H}\big\}\! =\! \sigma_0^2\bm{I}$. Then the received
 signal vector at the $k$th node,  where $1\! \le\! k\! \le\! K$, can be expressed as
\begin{align}\label{eq1}
 \bm{x}_k =& \bm{H}_k \bm{F}_k \bm{x}_{k-1} + \bm{n}_k ,
\end{align}
 where $\bm{H}_k$ is the $k$th hop channel matrix, $\bm{x}_{k-1}$ is the transmitted
 signal vector from the preceding node, and $\bm{n}_k$ is the additive white Gaussian
 noise (AWGN) vector at the $k$th node with the covariance matrix
 $\sigma_{\bm{n}_k}^2\bm{I}$, while the forward matrix $\bm{F}_k$
 satisfies the following hybrid structure
\begin{align}\label{eq2}
 \bm{F}_k =& \bm{F}_{{\rm AL},k} \bm{F}_{{\rm D},k} \bm{F}_{{\rm AR},k} ,
\end{align}
 in which $\bm{F}_{{\rm AL},k}$, $\bm{F}_{{\rm D},k}$, and $\bm{F}_{{\rm AR},k}$
 are the analog transmit precoder matrix, digital forward matrix, and analog
 receive combiner matrix for the $k$th hop or the $(k\! -\! 1)$th node, respectively.
 In particular, $\bm{F}_{{\rm AR},1}\! =\! \bm{I}$.
	
 Owing to the time varying nature and limited training resource, the channel
 state information (CSI) available at a node is imperfect. Hence, we model the
 channel matrix $\bm{H}_k$ by
\begin{align}\label{eq3}
 \bm{H}_k =& \widehat{\bm{H}}_k + \bm{H}_{{\rm W},k}\bm{\Psi}_k^{\frac{1}{2}} ,
\end{align}
 where $\widehat{\bm{H}}_k$ is the estimated channel matrix available, and the
 elements of $\bm{H}_{{\rm W},k}$ are i.i.d. random variables with zero mean and
 unit power. The positive semidefinite matrix $\bm{\Psi}_k$ is the transmit
 correlation matrix of the channel errors. The detailed derivation of $\bm{\Psi}_k$
 is beyond the scope of this paper and readers are recommended to referred to
 \cite{XingTSP2013}. Basically, $\bm{\Psi}_k$ is a function of training sequence.
	
 At the destination, i.e., node $K$, the desired signal $\bm{x}_0$ may be recovered
 from the noise corrupted observation $\bm{x}_K$ via a hybrid linear equalizer, which
 can be expressed as
\begin{align}\label{eq4}
 \widehat{\bm{x}}_0 =& \bm{G}_{\rm D} \bm{G}_{\rm A} \bm{x}_K ,
\end{align}
 where $\bm{G}_{\rm D}$ and $\bm{G}_{\rm A}$ denote the digital and analog equalizers
 of the hybrid transceiver at the destination, respectively. Given the hybrid linear
 equalizer and all the forward matrices $\{\bm{F}_k\}_{k=1}^K$, the corresponding mean
 squared error (MSE) matrix is defined by \cite{XingTVT2017}
\begin{align}\label{eq5}
 & \bm{\Phi}_{\rm MSE}^{\rm L}\big(\bm{G}_{\rm D},\bm{G}_{\rm A},\{\bm{F}_k\}_{k=1}^K\big)
 = \mathbb{E}\big\{\big(\bm{G}_{\rm D}\bm{G}_{\rm A}\bm{x}_K - \bm{x}_0\big)
  \big(\bm{G}_{\rm D}\bm{G}_{\rm A}\bm{x}_K - \bm{x}_0\big)^{\rm H}\big\} .
\end{align}
 As there is no constraint for the digital equalizer, the optimal $\bm{G}_{\rm D}$ can be
 derived in closed form \cite{XingTSP2013}. Substituting this optimal $\bm{G}_{\rm D}$
 into (\ref{eq5}), the data estimation MSE matrix can be expressed as
\begin{align}\label{eq6}
 & \bm{\Phi}_{\rm MSE}^{\rm L}\big(\bm{G}_{\rm A},\{\bm{F}_k\}_{k=1}^K\big) =
  \sigma_0^2\bm{I} - \sigma_0^4\big(\bm{G}_{\rm A}\widehat{\bm{H}}_K\bm{F}_K
  \widehat{\bm{H}}_{K-1}\bm{F}_{K-1}\cdots \widehat{\bm{H}}_1\bm{F}_1\big)^{\rm H} \nonumber \\
 & \hspace{10mm} \times \big(\bm{G}_{\rm A}\widehat{\bm{H}}_K\bm{F}_K\bm{R}_{\bm{x}_{K-1}}
  \bm{F}_K^{\rm H}\widehat{\bm{H}}_K^{\rm H}\bm{G}_{\rm A}^{\rm H} + \bm{K}_{\bm{n}_K}\big)^{-1}
  \big(\bm{G}_{\rm A}\widehat{\bm{H}}_K\bm{F}_K\widehat{\bm{H}}_{K-1}\bm{F}_{K-1}
  \cdots \widehat{\bm{H}}_1\bm{F}_1\big) ,
\end{align}
 where $\bm{K}_{\bm{n}_K}$ is the equivalent noise covariance matrix at destination,
 which can be expressed as
\begin{align}\label{eq7}
 \bm{K}_{\bm{n}_K} =& \bm{G}_{\rm A}\bm{R}_{\bm{n}_K}\bm{G}_{\rm A}^{\rm H} +
  {\rm Tr}\big(\bm{F}_K\bm{R}_{\bm{x}_{K-1}}\bm{F}_K^{\rm H}\bm{\Psi}_K\big)\bm{G}_{\rm A}
  \bm{G}_{\rm A}^{\rm H} ,
\end{align}
 while the covariance matrix $\bm{R}_{\bm{x}_k}$ of $\bm{x}_k$, for $ 1 < k \le K $, is
 given by
\begin{align}\label{eq8}
 \bm{R}_{\bm{x}_k} =& \widehat{\bm{H}}_k\bm{F}_k\bm{R}_{\bm{x}_{k-1}}\bm{F}_k^{\rm H}
  \widehat{\bm{H}}_k^{\rm H} + \bm{K}_{\bm{n}_k} ,
\end{align}
 in which
\begin{align}\label{eq9}
 \bm{K}_{\bm{n}_k} =& \sigma_{\bm{n}_k}^2\bm{I} + {\rm Tr}\big(\bm{F}_k\bm{R}_{\bm{x}_{k-1}}
  \bm{F}_k^{\rm H}\bm{\Psi}_k\big)\bm{I} .
\end{align}
 Note that $\bm{R}_{\bm{x}_0}=\sigma_0^2\bm{I}$.
	
 Based on the hybrid linear data estimation, nonlinear transceivers can further
 be implemented, for example, by using the THP  at the source or adopting the DFE at the
 destination. Let the lower triangular matrix $\bm{B}$ be the feedback matrix adopted in
 the THP or DFE. Then the corresponding data estimation MSE matrix can be expressed as
\begin{align}\label{eq10}
 \bm{\Phi}_{\rm MSE}^{\rm NL}\big(\bm{B},\bm{G}_{\rm A},\{\bm{F}_k\}_{k=1}^K\big)
  =& (\bm{I} + \bm{B})\bm{\Phi}_{\rm MSE}^{\rm L}\big(\bm{G}_{\rm A},\{\bm{F}_k\}_{k=1}^K\big)
  (\bm{I} + \bm{B})^{\rm H} .
\end{align}
	
 Based on the data estimation MSE matrices (\ref{eq6}) and (\ref{eq10}) for linear
 and nonlinear transceivers, respectively, the following hybrid transceiver
 optimization problems can be formulated. Specifically, the linear hybrid transceiver
 optimization for multi-hop communications can be formulated as
\begin{align}\label{eq11}
\begin{array}{cl}
 \min\limits_{\bm{G}_{\rm A},\{\bm{F}_k\}_{k=1}^K}\!\!\! & f_{\rm L}\big(\bm{d}\big[\bm{\Phi}_{\rm MSE}^{\rm L}
  (\bm{G}_{\rm A},\{\bm{F}_k\}_{k=1}^K)\big]\big) , \\
 {\rm{s.t.}}\!\!\! & {\rm Tr}\big(\bm{F}_k\bm{R}_{\bm{x}_{k-1}}\bm{F}_k^{\rm H})\le P_k , \
 \bm{F}_{{\rm AL},k} \in \mathcal{F}_{{\rm PL},k}, \bm{F}_{{\rm AR},k}\in \mathcal{F}_{{\rm PR},k},
  \bm{G}_{\rm A} \in \mathcal{F}_{\rm G} ,\!
\end{array}
\end{align}
 where $P_k$ is the maximum transmit power at the $k$th node, while $\mathcal{F}_{{\rm PL},k}$,
 $\mathcal{F}_{{\rm PR},k}$ and $\mathcal{F}_{\rm G}$ denote the corresponding analog matrix
 sets with proper dimensions and the elements of any matrix in these sets have constant
 amplitude.  The objective function $f_{\rm L}(\cdot )$ can be an additively Schur-convex
 or additively Schur-concave function of the diagonal elements of the data estimation MSE
 matrix $\bm{\Phi}_{\rm MSE}^{\rm L}\big(\bm{G}_{\rm A},\{\bm{F}_k\}_{k=1}^K\big)$ \cite{Majorization,XingTSP201502}.
 Similarly, the nonlinear hybrid transceiver optimization for multi-hop communications can
 be expressed as
\begin{align}\label{eq12}
\begin{array}{cl}
 \min\limits_{\bm{G}_{\rm A},\{\bm{F}_k\}_{k=1}^K} \!\!\!\! & f_{\rm NL}\big(\bm{d}\big[(\bm{I}\!
  +\! \bm{B}) \bm{\Phi}_{\rm MSE}^{\rm L}(\bm{G}_{\rm A},\{\bm{F}_k\}_{k=1}^K)(\bm{I}\! +\!
  \bm{B})^{\rm H} \big]\big) ,  \\
 {\rm{s.t.}}\!\!\!\! & {\rm Tr}\big(\bm{F}_k\bm{R}_{\bm{x}_{k-1}}\bm{F}_k^{\rm H})\le P_k ,
  \ \bm{F}_{{\rm AL},k} \in \mathcal{F}_{{\rm PL},k}, \bm{F}_{{\rm AR},k}\in \mathcal{F}_{{\rm PR},k},
  \bm{G}_{\rm A} \in \mathcal{F}_{\rm G},
\end{array}\end{align}
 where the objective function $f_{\rm NL}(\cdot )$ is a multiplicatively Schur-convex
 or multiplicatively Schur-concave function of the diagonal elements of
 $\bm{\Phi}_{\rm MSE}^{\rm NL}\big(\bm{B},\bm{G}_{\rm A}, \{\bm{F}_k\}_{k=1}^K\big)$ \cite{Majorization,Jorswieck07}.

\section{Problem Reformulation}\label{S3}
	
 To simplify the derivations for transceiver designs, we introduce the auxiliary variables
\begin{align}
 \bar{\bm{F}}_1 =& \bm{F}_1 \bm{Q}_0^{\rm H} , \label{eq13} \\
 \bar{\bm{F}}_k =& \bm{F}_k \bm{K}_{\bm{n}_{k-1}}^{\frac{1}{2}} \bm{\Sigma}_{k-1}^{\frac{1}{2}}
  \bm{Q}_{k-1}^{\rm H} , ~ 2\le k\le K , \label{eq14}
\end{align}
 where $\bm{Q}_k$ for $0\! \le\! k\! \le\! K-1$ are unitary matrices with proper dimensions,
 and for $2\! \le\! k\! \le\! K$,
\begin{align}\label{eq15}
 \bm{\Sigma}_{k\! -\! 1}\! =& \bm{K}_{\bm{n}_{k\! -\! 1}}^{-\frac{1}{2}} \widehat{\bm{H}}_{k\! -\! 1}\bm{F}_{k\! -\! 1}
  \bm{R}_{\bm{x}_{k\! -\! 2}}\bm{F}_{k\! -\! 1}^{\rm H}\widehat{\bm{H}}_{k\! -\! 1}^{\rm H}
  \bm{K}_{\bm{n}_{k\! -\! 1}}^{-\frac{1}{2}} \! +\! \bm{I} .\!
\end{align}
 Therefore, the linear data estimation MSE matrix can be reformulated as
\begin{align}\label{eq16}
 \bm{\Phi}_{\rm MSE}^{\rm L}\big(\bm{G}_{\rm A},\{\bar{\bm{F}}_k\}_{k=1}^K, \{\bm{Q}_k\}_{k=0}^{K-1}\big)
  = \sigma_0^2\bm{I} - \sigma_0^4 \bm{\Upsilon}^{\rm H} \bm{\Upsilon} ,
\end{align}
 where
\begin{align}\label{eq17}
 \bm{\Upsilon} =& \Big(\bm{\Sigma}_K^{-\frac{1}{2}} \bm{K}_{\bm{n}_K}^{-\frac{1}{2}} \bm{G}_{\rm A}
  \widehat{\bm{H}}_K \bar{\bm{F}}_K \bm{Q}_{K\! -\! 1} \bm{\Sigma}_{K\! -\! 1}^{-\frac{1}{2}}
  \bm{K}_{\bm{n}_{K\! -\! 1}}^{-\frac{1}{2}} \widehat{\bm{H}}_{K\! -\! 1} \bar{\bm{F}}_{K\! -\!1}
  \cdots \bm{Q}_1 \bm{\Sigma}_1^{-\frac{1}{2}} \bm{K}_{\bm{n}_1}^{-\frac{1}{2}} \widehat{\bm{H}}_1
  \bar{\bm{F}}_1 \bm{Q}_0\Big) ,
\end{align}
 in which
\begin{align}\label{eq18}
 \bm{\Sigma}_K =& \bm{K}_{\bm{n}_K}^{-\frac{1}{2}} \bm{G}_{\rm A} \widehat{\bm{H}}_K \bar{\bm{F}}_K
  \bm{R}_{\bm{x}_{K-1}} \bar{\bm{F}}_K^{\rm H} \widehat{\bm{H}}_K^{\rm H} \bm{G}_{\rm A}^{\rm H}
  \bm{K}_{\bm{n}_K}^{-\frac{1}{2}} + \bm{I} .
\end{align}

 Based on the reformulated data estimation matrix $\bm{\Phi}_{\rm MSE}^{\rm L}\big(\bm{G}_{\rm A},
 \{\bar{\bm{F}}_k\}_{k=1}^0,\{\bm{Q}_k\}_{k=0}^{K-1}\big)$, the linear transceiver
 optimization problem (\ref{eq11}) can be re-expressed as
\begin{align}\label{eq19}
\begin{array}{cl}
 \min\limits_{\bm{G}_{\rm A},\{\bar{\bm{F}}_k\},\{\bm{Q}_k\}}\ \ &
  f_{\rm L}\big(\bm{d}\big[\bm{\Phi}_{\rm MSE}^{\rm L} (\bm{G}_{\rm A},\{\bar{\bm{F}}_k\},
  \{\bm{Q}_k\})\big]\big) ,  \\
 {\rm{s.t.}} \ \ &{\rm Tr}\big(\bar{\bm{F}}_k\bar{\bm{F}}_k^{\rm H})\le P_k ,  \  \bm{F}_{{\rm AL},k}\! \in\! \mathcal{F}_{{\rm PL},k},
  \bm{F}_{{\rm AR},k}\! \in\! \mathcal{F}_{{\rm PR},k},
  \bm{G}_{\rm A}\! \in\! \mathcal{F}_{\rm G},
\end{array}
\end{align}
 where for notational simplification, we have dropped the ranges of $\{\bar{\bm{F}}_k\}$
 and $\{\bm{Q}_k\}$.

 Similarly, the nonlinear transceiver optimization problem (\ref{eq12}) can be
 rewritten in the following form
\begin{align}\label{eq20}
\begin{array}{cl}
 \min\limits_{\bm{G}_{\rm A},\{\bar{\bm{F}}_k\},\{\bm{Q}_k\}} \!\! &
  f_{\rm NL}\big(\bm{d}\big[(\bm{I}\! +\! \bm{B})
  \bm{\Phi}_{\rm MSE}^{\rm L}(\bm{G}_{\rm A},\{\bar{\bm{F}}_k\},\{\bm{Q}_k\})
  (\bm{I}\! +\! \bm{B})^{\rm H}\big]\big) , \\
 {\rm{s.t.}}\!\! & {\rm Tr}\big(\bar{\bm{F}}_k\bar{\bm{F}}_k^{\rm H})\le P_k , \ 
  \bm{F}_{{\rm AL},k}\! \in\! \mathcal{F}_{{\rm PL},k},
  \bm{F}_{{\rm AR},k}\! \in\! \mathcal{F}_{{\rm PR},k},
  \bm{G}_{\rm A}\! \in\! \mathcal{F}_{\rm G}.\!
\end{array}
\end{align}
 The optimal lower triangular matrix $\bm{B}$ satisfies \cite{XingJSAC2012,XingTSP2016}
\begin{align}\label{eq21}
 \bm{I} + \bm{B}_{\rm opt} =& {\rm diag}\{\bm{d}[\bm{L}]\}\bm{L}^{-1} ,
\end{align}
 where $\bm{L}$ is the lower triangular matrix of the following Cholesky decomposition
\begin{align}\label{eq22}
 \bm{\Phi}_{\rm MSE}^{\rm L}\big(\bm{G}_{\rm A},\{\bar{\bm{F}}_k\},\{\bm{Q}_k\}\big)
  =& \bm{L}\bm{L}^{\rm H} .
\end{align}
 As a result, the general nonlinear transceiver optimization problem (\ref{eq20})
 can be rewritten as
\begin{align}\label{eq23}
\begin{array}{cl}
 \min\limits_{\bm{G}_{\rm A},\{\bar{\bm{F}}_k\},\{\bm{Q}_k\}}\ \ & f_{\rm NL}\big(\bm{d}^2[\bm{L}]\big) ,  \\
 {\rm{s.t.}}\!\! & \bm{\Phi}_{\rm MSE}^{\rm L}\big(\bm{G}_{\rm A},\{\bar{\bm{F}}_k\},\{\bm{Q}_k\}\big)
   = \bm{L}\bm{L}^{\rm H} , \
  {\rm Tr}\big(\bar{\bm{F}}_k\bar{\bm{F}}_k^{\rm H})\le P_k , \\
 \!\! &  \bm{F}_{{\rm AL},k}\! \in\! \mathcal{F}_{{\rm PL},k},
  \bm{F}_{{\rm AR},k}\! \in\! \mathcal{F}_{{\rm PR},k},
  \bm{G}_{\rm A}\! \in\! \mathcal{F}_{\rm G} .\!
\end{array}
\end{align}

 In the following sections, it is shown that the optimal $\{\bm{Q}_k\}$, $\{\bar{\bm{F}}_k\}$
 and $\bm{G}_{\rm A}$ can be derived separately for both linear and nonlinear transceiver
 designs of the multi-hop AF MIMO relay system with different objective functions.
		
\section{Optimal Unitary Matrices}\label{S4}

 Since $\{\bm{Q}_k\}$ do not appear in the constraints, based on our previous works
 \cite{XingTSP201502,XingTSP2016}, we can easily derive the optimal $\bm{Q}_k$ for
 $1\! \le\! k\! \le\! K-1$, as summarized in the following conclusion.

\begin{conclusion}\label{C1}
 Define the following singular value decompositions (SVDs)
\begin{align}
 & \bm{\Sigma}_k^{-\frac{1}{2}} \bm{K}_{\bm{n}_k}^{-\frac{1}{2}} \widehat{\bm{H}}_k \bar{\bm{F}}_k
  = \bm{U}_k \bm{\Lambda}_k \bm{V}_k^{\rm H} , ~  1\le k < K , \label{eq24} \\
 & \bm{\Sigma}_K^{-\frac{1}{2}} \bm{K}_{\bm{n}_K}^{-\frac{1}{2}} \bm{G}_{\rm A} \widehat{\bm{H}}_K
  \bar{\bm{F}}_K = \bm{U}_K \bm{\Lambda}_K \bm{V}_K^{\rm H} . \label{eq25}
\end{align}
 Then the optimal $\bm{Q}_k$ for $1\le k\le K-1$ are given by
\begin{align}\label{eq26}
 \bm{Q}_{k,{\rm opt}} =& \bm{V}_{k+1} \bm{U}_k^{\rm H} .
\end{align}
\end{conclusion}

 The optimal $\bm{Q}_0$ depends on the objective function, and it is discussed case by case.
	
\subsection{Linear Transceiver Designs}\label{S4.1}

 Consider the additively Schur-convex objective function for $f_{\rm L}(\cdot )$, namely,
\begin{align}\label{eq27}
 & \textbf{Obj.1}: f_{\text{A-Schur}}^{\text{Convex}}\left(\bm{d}\big[\bm{\Phi}_{\rm MSE}^{\rm L}
  (\bm{G}_{\rm A},\{\bar{\bm{F}}_k\},\{\bm{Q}_k\})\big]\right) .
\end{align}
 Then according to \cite{XingTSP201501},
\begin{align}\label{eq28}
  \bm{Q}_{0,\text{Opt}} =& \bm{V}_1 \bar{\bm{U}}_{\text{DFT}}^{\rm H} ,
\end{align}
 where the unitary matrix $\bar{\bm{U}}_{\text{DFT}}$ is the discrete Fourier transform
 (DFT) matrix of appropriate dimension, which ensures that all the diagonal elements
 of the data estimation MSE matrix are identical. On the other hand, when the
 objective function is additively Schur-concave, that is,
\begin{align}\label{eq29}
 & \textbf{Obj.2}: f_{{\text{A-Schur}}}^{\text{Concave}}\left(\bm{d}\big[\bm{\Phi}_{\rm MSE}^{\rm L}
  (\bm{G}_{\rm A},\{\bar{\bm{F}}_k\},\{\bm{Q}_k\})\big]\right) ,
\end{align}
 we have \cite{XingTSP201501}
\begin{align}\label{eq30}
 \bm{Q}_{0,\text{opt}} =& \bm{V}_1 .
\end{align}
 It can be seen that with the additively Schur-concave objective function, the matrix
 version of the signal-to-noise ratio (SNR) is a diagonal matrix at the optimal solution of
 $\bm{Q}_{0,\text{opt}}$.
	
 Based on the optimal $\big\{\bm{Q}_{k,\text{opt}}\big\}_{k=0}^{K\! -\! 1}$, the linear
 transceiver optimization problem (\ref{eq19}) becomes
\begin{align}\label{eq31}
\begin{array}{cl}
 \min\limits_{\bm{G}_{\rm A},\{\bar{\bm{F}}_k\}}\!\! & f_{\rm L}\Big(\Big\{\bm{\lambda}\big\{
  \bar{\bm{F}}_k^{\rm H} \widehat{\bm{H}}_k^{\rm H} \bm{K}_{\bm{n}_k}^{-1} \widehat{\bm{H}}_k
  \bar{\bm{F}}_k\big\} \Big\} \Big) , \\
 {\rm{s.t.}}\!\! & {\rm Tr}\big(\bar{\bm{F}}_k\bar{\bm{F}}_k^{\rm H})\le P_k ,
  \ \bm{F}_{{\rm AL},k} \in \mathcal{F}_{{\rm PL},k}, \bm{F}_{{\rm AR},k}\in \mathcal{F}_{{\rm PR},k},
  \bm{G}_{\rm A} \in \mathcal{F}_{\rm G} ,\!
\end{array}
\end{align}
 where again for notational simplification, we have dropped the range of
 $\Big\{\bm{\lambda}\big\{ \bar{\bm{F}}_k^{\rm H}\widehat{\bm{H}}_k^{\rm H} \bm{K}_{\bm{n}_k}^{-1}
 \widehat{\bm{H}}_k \bar{\bm{F}}_k\big\} \Big\}$.

\subsection{Nonlinear Transceiver Designs}\label{S4.2}

 For the nonlinear transceiver designs with THP or DFE, when the objective
 $f_{\rm NL}(\cdot )$ is multiplicatively Schur-convex w.r.t. the diagonal elements of
 the data estimation MSE matrix, namely,
\begin{align}\label{eq32}
 \textbf{Obj.3}: & f_{{\text{M-Schur}}}^{\text{Convex}}\left(\bm{d}^2[\bm{L}]\right) , \ 
 {\rm{with}} \ \bm{\Phi}_{\rm MSE}^{\rm L}\big(\bm{G}_{\rm A},\{\bar{\bm{F}}_k\},\{\bm{Q}_k\}\big)
   = \bm{L}\bm{L}^{\rm H} ,
\end{align}
 the optimal solution of $\bm{Q}_0$ is given by \cite{XingTSP201501}
\begin{align}\label{eq33}
 \bm{Q}_{0,\text{opt}} =& \bm{V}_1 \bar{\bm{U}}_{\text{GMD}}^{\rm H} ,
\end{align}
 where the unitary matrix $\bar{\bm{U}}_{\text{GMD}}$ makes sure that the lower
 triangular matrix $\bm{L}$ has the same diagonal elements. On the other hand, when
 the objective function is multiplicatively Schur-concave w.r.t. the diagonal
 elements of the data estimation MSE matrix, i.e.,
\begin{align}\label{eq34}
 \textbf{Obj.4}: & f_{{\text{M-Schur}}}^{\text{Concave}}\left(\bm{d}^2[\bm{L}]\right) , \
 {\rm{with}} \ \bm{\Phi}_{\rm MSE}^{\rm L}\big(\bm{G}_{\rm A},\{\bar{\bm{F}}_k\},\{\bm{Q}_k\}\big)
   = \bm{L}\bm{L}^{\rm H} ,
\end{align}
 the optimal solution of $\bm{Q}_0$ is given by \cite{XingTSP201501}
\begin{align}\label{eq35}
 \bm{Q}_{0,\text{opt}} =& \bm{V}_1 .
\end{align}
 It is obvious that when the objective function is multiplicatively Schur-concave,
 the matrix version SNR is a diagonal matrix at the optimal solution of $\bm{Q}_{0,{\rm opt}}$.
		
 Based on the optimal solution of $\big\{\bm{Q}_{k,\text{opt}}\big\}_{k=0}^{K-1}$, the
 nonlinear transceiver optimization problem can be rewritten as
\begin{align}\label{eq36}
\begin{array}{cl}
 \min\limits_{\bm{G}_{\rm A},\{\bar{\bm{F}}_k\}} \ & f_{\rm NL}\Big(\Big\{\bm{\lambda}\big\{
  \bar{\bm{F}}_k^{\rm H} \widehat{\bm{H}}_k^{\rm H} \bm{K}_{\bm{n}_k}^{-1} \widehat{\bm{H}}_k
  \bar{\bm{F}}_k\big\} \Big\} \Big) , \\
 {\rm{s.t.}} \ & {\rm Tr}\big(\bar{\bm{F}}_k\bar{\bm{F}}_k^{\rm H}\big)\le P_k ,
  \ \bm{F}_{{\rm AL},k} \in \mathcal{F}_{{\rm PL},k}, \bm{F}_{{\rm AR},k}\in \mathcal{F}_{{\rm PR},k},
  \bm{G}_{\rm A} \in \mathcal{F}_{\rm G} .
\end{array}
\end{align}
	
 In a nutshell, for linear transceiver optimization and nonlinear transceiver
 optimization, the optimal solution is a Pareto optimal solution of the following
 optimization problem
\begin{align}\label{Vector_Opt} 
\begin{array}{cl}\!\!\!
 \max\limits_{\bm{G}_{\rm A},\{\bar{\bm{F}}_k\}} & \Big\{\bm{\lambda}\big\{
  \bar{\bm{F}}_k^{\rm H} \widehat{\bm{H}}_k^{\rm H} \bm{K}_{\bm{n}_k}^{-1} \widehat{\bm{H}}_k
  \bar{\bm{F}}_k\big\} \Big\} ,\\
 {\rm{s.t.}} & {\rm Tr}\big(\bar{\bm{F}}_k\bar{\bm{F}}_k^{\rm H}\big)\le P_k, \
  \bm{F}_{{\rm AL},k} \in \mathcal{F}_{{\rm PL},k}, \bm{F}_{{\rm AR},k}\in \mathcal{F}_{{\rm PR},k},
  \bm{G}_{\rm A} \in \mathcal{F}_{\rm G} .\!
\end{array}\!
\end{align}
 Therefore, the common structures of all the Pareto optimal solutions of this vector
 optimization problem are the structures of the optimal solutions of our linear
 transceiver optimization problem and nonlinear transceiver optimization problem. In
 the following, we will derive the optimal structures of the Pareto optimal solutions.
 Since for multi-hop AF MIMO communications, the hybrid transceiver optimizations are
 different in the first hop, the intermediate hops, and the final hop, we will
 investigate these hybrid transceiver optimizations case by case.
	
\section{Optimal Structures of Hybrid Transceivers}\label{S5}
	
\subsection{First Hop}\label{S5.1}

 The first-hop communication occurs between the source, node 1, and the first relay,
 node 2. By defining
\begin{align}\label{eq38}
 \bar{\bm{F}}_{{\rm D},1} =& \bm{ F}_{{\rm D},1} \bm{R}_{\bm{x}_0}^{\frac{1}{2}} ,
\end{align}
 the vector optimization problem (\ref{Vector_Opt}) for the first hop can be expressed
 in the following form
\begin{align}\label{E_Opt_First} 
\begin{array}{cl}
 \max\limits_{\bar{\bm{F}}_1} \!\! & \bm{\lambda}\Big\{\bar{\bm{F}}_{{\rm D},1}^{\rm H}
  \bm{F}_{{\rm AL},1}^{\rm H} \widehat{\bm{H}}_1^{\rm H} \bm{K}_{\bm{n}_1}^{-1}
  \widehat{\bm{H}}_1 \bm{F}_{{\rm AL},1} \bar{\bm{F}}_{{\rm D},1}\Big\} , \\
 {\rm{s.t.}} \!\! & {\rm Tr}\Big( \bm{F}_{{\rm AL},1} \bar{\bm{F}}_{{\rm D},1}
  \bar{\bm{F}}_{{\rm D},1}^{\rm H} \bm{F}_{{\rm AL},1}^{\rm H}\Big)\le P_1 , 
  \ \bm{F}_{{\rm AL},1} \in \mathcal{F}_{{\rm PL},1} .
\end{array}
\end{align}
 Noting the equivalent noise covariance matrix in the first hop
\begin{align}\label{eq40}
 \bm{K}_{\bm{n}_1}\! =& \Big( \sigma_{\bm{n}_1}^2\! +\! {\rm{Tr}}\big( \bm{F}_{{\rm AL},1}
  \bar{\bm{F}}_{{\rm D},1} \bar{\bm{F}}_{{\rm D},1}^{\rm H} \bm{F}_{{\rm AL},1}^{\rm H}
  \bm{\Psi}_1\big) \Big) \bm{I} \triangleq \eta_1 \bm{I},
\end{align}
 it is obvious that the forward matrix optimization in the first hop is challenging to
 solve, and some reformulations are needed.

 Note that the following power constraint
\begin{align}\label{eq41}
 {\rm Tr}\big(\bm{F}_{{\rm AL},1} \bar{\bm{F}}_{{\rm D},1} \bar{\bm{F}}_{{\rm D},1}^{\rm H}
  \bm{F}_{{\rm AL},1}^{\rm H}\big)\le P_1
\end{align}
 is equivalent to the following one
\begin{align}\label{eq42}
 \frac{{\rm Tr}\Big(\big( \sigma_{\bm{n}_1}^2 \bm{I} + P_1 \bm{\Psi}_1\big) \bm{F}_{{\rm AL},1}
  \bar{\bm{F}}_{{\rm D},1} \bar{\bm{F}}_{{\rm D},1}^{\rm H} \bm{F}_{{\rm AL},1}^{\rm H}\Big)}{\eta_1}\le P_1 .
\end{align}
 Hence the optimization problem (\ref{E_Opt_First}) is equivalent to
\begin{align}\label{E_Opt_First_A} 
\begin{array}{cl}
 \max\limits_{\bar{\bm{F}}_1} \!\! & \bm{\lambda}\Big\{\bar{\bm{F}}_{{\rm D},1}^{\rm H}
  \bm{F}_{{\rm AL},1}^{\rm H} \widehat{\bm{H}}_1^{\rm H} \bm{K}_{\bm{n}_1}^{-1}
  \widehat{\bm{H}}_1 \bm{F}_{{\rm AL},1} \bar{\bm{F}}_{{\rm D},1}\Big\} , \\
 {\rm{s.t.}} \!\! & \frac{{\rm Tr}\Big(\big( \sigma_{\bm{n}_1}^2 \bm{I} + P_1 \bm{\Psi}_1\big) \bm{F}_{{\rm AL},1}
  \bar{\bm{F}}_{{\rm D},1} \bar{\bm{F}}_{{\rm D},1}^{\rm H} \bm{F}_{{\rm AL},1}^{\rm H}\Big)}{\eta_1}\le P_1 , \
 \bm{F}_{{\rm AL},1} \in \mathcal{F}_{{\rm PL},1} .\!
\end{array}
\end{align}
 By defining the following auxiliary variables
\begin{align}
 & \widetilde{\bm{F}}_{{\rm D},1}\! =\! \eta_1^{-\frac{1}{2}} \Big( \bm{F}_{{\rm AL},1}^{\rm H}
  \big(\sigma_{\bm{n}_1}^2 \bm{I} + P_1 \bm{\Psi}_1\big) \bm{F}_{{\rm AL},1}\Big)^{\frac{1}{2}}
  \bar{\bm{F}}_{{\rm D},1} , \label{eq44} \\
 & \bm{\Pi}_1\! =\! \big( \sigma_{\bm{n}_1}^2 \bm{I}\! +\! P_1 \bm{\Psi}_1\big)^{\frac{1}{2}} \bm{F}_{{\rm AL},1}
  \big( \bm{F}_{{\rm AL},1}^{\rm H} \big(\sigma_{\bm{n}_1}^2 \bm{I}\! +\! P_1 \bm{\Psi}_1\big)
  \bm{F}_{{\rm AL},1}\big)^{-\frac{1}{2}}\! , \label{eq45}
\end{align}
 the vector optimization problem (\ref{E_Opt_First_A}) can be rewritten in the following form
\begin{align}\label{E_Opt_First_B} 
\begin{array}{cl}
 \max\limits_{\bar{\bm{F}}_1} \!\! & \bm{\lambda}\Big\{\widetilde{\bm{F}}_{{\rm D},1}^{\rm H}
  \bm{\Pi}_1^{\rm H} \big(\sigma_{\bm{n}_1}\bm{I} + P_1\bm{\Psi}_1\big)^{-\frac{1}{2}}
   \widehat{\bm{H}}_1^{\rm H}  \widehat{\bm{H}}_1 \big(\sigma_{\bm{n}_1}\bm{I} + P_1\bm{\Psi}_1\big)^{-\frac{1}{2}}
   \bm{\Pi}_1\widetilde{\bm{F}}_{{\rm D},1} \Big\} , \\
 {\rm{s.t.}} \!\! &  {\rm{Tr}}\big(\widetilde{\bm{F}}_{{\rm D},1}\widetilde{\bm{F}}_{{\rm D},1}^{\rm H}
  \big)\le P_1 , \  \bm{F}_{{\rm AL},1} \in \mathcal{F}_{{\rm PL},1} , \!
\end{array}
\end{align}
 which is equivalent to the following matrix-monotonic optimization problem
\begin{align}\label{MM_first_hop} 
\begin{array}{cl}
 \max\limits_{\bar{\bm{F}}_1} \!\! & \widetilde{\bm{F}}_{{\rm D},1}^{\rm H} \bm{\Pi}_1^{\rm H}
  \big(\sigma_{\bm{n}_1}\bm{I} + P_1\bm{\Psi}_1\big)^{-\frac{1}{2}}\widehat{\bm{H}}_1^{\rm H}
  \widehat{\bm{H}}_1\big(\sigma_{\bm{n}_1}\bm{I} + P_1\bm{\Psi}_1\big)^{-\frac{1}{2}}
  \bm{\Pi}_1\widetilde{\bm{F}}_{{\rm D},1} , \\
 {\rm{s.t.}}\!\! &  {\rm{Tr}}\big(\widetilde{\bm{F}}_{{\rm D},1}\widetilde{\bm{F}}_{{\rm D},1}^{\rm H}
  \big)\le P_1 , \  \bm{F}_{{\rm AL},1} \in \mathcal{F}_{{\rm PL},1}.\!
\end{array}
\end{align}
	
 From (\ref{eq45}), it is obvious that $\bm{\Pi}_1$ is determined by the singular matrices of
 analog transmit precoder $\bm{F}_{{\rm AL},1}$ and the nonzero singular values of $\bm{\Pi}_1$
 are all ones. In other words, we only need to analyze the SVD unitary matrices of $\bm{\Pi}_1$.
 Furthermore, in the optimization problem (\ref{MM_first_hop}), the constraint is unitary invariant
 to the digital forward matrix $\widetilde{\bm{F}}_{{\rm D},1}$. This means that we only need to
 analyze the left SVD unitary matrix of $\bm{\Pi}_1$, which is equivalent to the left SVD unitary
 matrix of $\big(\sigma_{\bm{n}_1}^2\bm{I} + P_1\bm{\Psi}_1\big)^{\frac{1}{2}}\bm{F}_{{\rm AL},1}$.
 Then the following conclusion obviously holds.

\begin{conclusion}\label{C2}
 The singular values of $\big(\sigma_{\bm{n}_1}^2\bm{I}+P_1\bm{\Psi}_1)^{\frac{1}{2}}
 \bm{F}_{{\rm AL},1}$ do not affect the system performance. The left eigenvectors of the
 SVD for $\big(\sigma_{\bm{n}_1}^2\bm{I}+P_1\bm{\Psi}_1)^{\frac{1}{2}} \bm{F}_{{\rm AL},1}$
 have the maximum inner product with respect to the eigenvectors $\bm{V}_{\bm{\mathcal{H}}_1}$,
 defined by the following SVD
\begin{align}\label{eq48}
 \widehat{\bm{H}}_1\big(\sigma_{\bm{n}_1}^2\bm{I}\! +\! P_1\bm{\Psi}_1\big)^{-\frac{1}{2}}\! =\!
  \bm{U}_{\bm{\mathcal{H}}_1}\bm{\Lambda}_{\bm{\mathcal{H}}_1}\bm{V}_{\bm{\mathcal{H}}_1}^{\rm H}
  \ \text{with} \ \bm{\Lambda}_{\bm{\mathcal{H}}_1} \searrow \! . \!
\end{align}
\end{conclusion}

 The optimal structure of $\widetilde{\bm{F}}_{{\rm D},1}$ is readily derived in the following
 conclusion \cite{XingTSP2013,XingTSP201502}.

\begin{conclusion}\label{C3}
 Based on the SVD
\begin{align}\label{eq49}
 \!\!\! \widehat{\bm{H}}_1\! \big(\sigma_{\bm{n}_1}^2\! \bm{I}\! +\! P_1\bm{\Psi}_1\big)^{-\frac{1}{2}}
  \bm{\Pi}_1\!\! =\! \bm{U}_{{\bm\Pi},1} \bm{\Lambda}_{{\bm\Pi},1}\bm{V}_{{\bm\Pi},1}^{\rm H}
  \, \text{with} \, \bm{\Lambda}_{{\bm\Pi},1}\! \searrow \!
\end{align}
 for given $\bm{F}_{{\rm AL},1}$, all the Pareto optimal $\widetilde{\bm{F}}_{{\rm D},1}$
 of the optimization problem (\ref{MM_first_hop}) satisfy the following structure
\begin{align}\label{eq50}
 \widetilde{\bm{F}}_{{\rm D},1} = \bm{V}_{{\bm\Pi},1} \bm{\Lambda}_{\widetilde{\bm{F}}_{{\rm D},1}}
  \bm{U}_{\rm Arb}^{\rm H} ,
\end{align}
 where $\bm{\Lambda}_{\widetilde{\bm{F}}_{{\rm D},1}}$ is a rectangular diagonal matrix, and
 $\bm{U}_{\rm Arb}$ is an arbitrary right unitary matrix with proper dimension.
\end{conclusion}

 Based on Conclusion~\ref{C3} and the definition (\ref{eq44}), when the optimal
 $\widetilde{\bm{F}}_{{\rm D},1}$ is given, the optimal $\bar{\bm{F}}_{{\rm D},1}$
 is readily computed as
\begin{align}\label{eq51}
 \bar{\bm{F}}_{{\rm D},1} =& \sqrt{\frac{P_1}{\alpha_1}} \Big( \bm{F}_{{\rm AL},1}^{\rm H}
  \big(\sigma_{\bm{n}_1}^2 \bm{I} + P_1 \bm{\Psi}_1\big) \bm{F}_{{\rm AL},1}\Big)^{-\frac{1}{2}}
  \widetilde{\bm{F}}_{{\rm D},1} ,
\end{align}
 in which $\alpha_1$ is given by
\begin{align}\label{eq52}
 \alpha_1\! =& {\rm{Tr}}\Big(\! \Big(\! \bm{F}_{{\rm AL},1}^{\rm H}\big(\sigma_{\bm{n}_1}^2 \bm{I}
  \! +\! P_1 \bm{\Psi}_1\big) \bm{F}_{{\rm AL},1}\Big)^{-\frac{1}{2}} \bm{F}_{{\rm AL},1}^{\rm H}
  \bm{F}_{{\rm AL},1} \Big(\bm{F}_{{\rm AL},1}^{\rm H}\big(\sigma_{\bm{n}_1}^2 \bm{I}
  \! +\! P_1 \bm{\Psi}_1\big) \bm{F}_{{\rm AL},1}\Big)^{-\frac{1}{2}} \widetilde{\bm{F}}_{{\rm D},1}
  \widetilde{\bm{F}}_{{\rm D},1}^{\rm H} \Big)\! .\!
\end{align}

\subsection{Intermediate Hops}\label{S5.2}

 First define
\begin{align}\label{eq53}
 \bar{\bm{F}}_{{\rm AR},k} = \bm{F}_{{\rm AR},k} \bm{R}_{\bm{x}_{k-1}}^{\frac{1}{2}} , \  2\le k\le K .
\end{align}
 Then the optimal forward matrices in the intermediate hops, namely, the hops $2\le k\le K-1$,
 are the Pareto optimal solutions of the following optimizations
\begin{align}\label{E_Opt_Medium} 
\begin{array}{cl}
 \max\limits_{\bar{\bm{F}}_k}\!\! &  \bm{\lambda}\Big\{ \bar{\bm{F}}_{{\rm AR},k}^{\rm H}
  \bm{F}_{{\rm D},k}^{\rm H} \bm{F}_{{\rm AL},k}^{\rm H} \widehat{\bm{H}}_k^{\rm H}
  \bm{K}_{\bm{n}_k}^{-1} \widehat{\bm{H}}_k \bm{F}_{{\rm AL},k} \bm{F}_{{\rm D},k}
  \bar{\bm{F}}_{{\rm AR},k} \Big\} , \\
 {\rm{s.t.}}\!\!  & {\rm{Tr}}\big( \bm{F}_{{\rm AL},k} \bm{F}_{{\rm D},k} \bar{\bm{F}}_{{\rm AR},k}
  \bar{\bm{F}}_{{\rm AR},k}^{\rm H} \bm{F}_{{\rm D},k}^{\rm H} \bm{F}_{{\rm AL},k}^{\rm H}\big)\le P_k ,
  \ \bm{F}_{{\rm AL},k}\in \mathcal{F}_{{\rm PL},k}, \ \bm{F}_{{\rm AR},k}\in \mathcal{F}_{{\rm PR},k}. \!
\end{array}
\end{align}
 Noting the equivalent noise covariance matrices
\begin{align}\label{eq55}
 \bm{K}_{\bm{n}_k}\! =& \Big(\! \sigma_{\bm{n}_k}^2\! +\! {\rm{Tr}}\Big( \bm{F}_{{\rm AL},k} \bm{F}_{{\rm D},k}
  \bar{\bm{F}}_{{\rm AR},k} \bar{\bm{F}}_{{\rm AR},k}^{\rm H} \bm{F}_{{\rm D},k}^{\rm H}
  \bm{F}_{{\rm AL},k}^{\rm H} \bm{\Psi}_k\Big) \! \Big) \bm{I}
 \triangleq  \eta_k \bm{I} ,
\end{align}
 the power constraints
\begin{align}\label{eq56}
 {\rm{Tr}}\big(\bm{F}_{{\rm AL},k}\bm{F}_{{\rm D},k}\bar{\bm{F}}_{{\rm AR},k}\bar{\bm{F}}_{{\rm AR},k}^{\rm H}
  \bm{F}_{{\rm D},k}^{\rm H} \bm{F}_{{\rm AL},k}^{\rm H}\big)\le P_k
\end{align}
 are equivalent to
\begin{align}\label{eq57}
 \frac{{\rm{Tr}}\Big(\! \big(\sigma_{\bm{n}_k}^2\! \bm{I}\! +\! P_k \bm{\Psi}_k\big)\bm{F}_{{\rm AL},k}
  \bm{F}_{{\rm D},k} \bar{\bm{F}}_{{\rm AR},k} \bar{\bm{F}}_{{\rm AR},k}^{\rm H}
  \bm{F}_{{\rm D},k}^{\rm H} \bm{F}_{{\rm AL},k}^{\rm H}\! \Big)}{\eta_k}\! \le\! P_k .
\end{align}
 As a result, after replacing the original constraint, the optimization problem
 (\ref{E_Opt_Medium}) is equivalent to
\begin{align}\label{E_Opt_Medium_A} 
\begin{array}{cl}
 \max\limits_{\bar{\bm{F}}_k} \!\! & \bm{\lambda}\Big\{ \bar{\bm{F}}_{{\rm AR},k}^{\rm H}
  \bm{F}_{{\rm D},k}^{\rm H} \bm{F}_{{\rm AL},k}^{\rm H} \widehat{\bm{H}}_k^{\rm H}
  \bm{K}_{\bm{n}_k}^{-1} \widehat{\bm{H}}_k \bm{F}_{{\rm AL},k} \bm{F}_{{\rm D},k}
  \bar{\bm{F}}_{{\rm AR},k} \Big\} ,  \\
 {\rm{s.t.}} \  & \frac{{\rm{Tr}}\Big(\! \big(\sigma_{\bm{n}_k}^2\! \bm{I}\! +\! P_k \bm{\Psi}_k\big)\bm{F}_{{\rm AL},k}
  \bm{F}_{{\rm D},k} \bar{\bm{F}}_{{\rm AR},k} \bar{\bm{F}}_{{\rm AR},k}^{\rm H}
  \bm{F}_{{\rm D},k}^{\rm H} \bm{F}_{{\rm AL},k}^{\rm H}\! \Big)}{\eta_k}\! \le\! P_k , \
   \bm{F}_{{\rm AL},k}\in \mathcal{F}_{{\rm PL},k}, \ \bm{F}_{{\rm AR},k}\in \mathcal{F}_{{\rm PR},k}. \!
\end{array}
\end{align}
 By defining the following auxiliary variables
\begin{align}
 \widetilde{\bm{F}}_{{\rm D},k} =& \eta_k^{-\frac{1}{2}} \Big( \bm{F}_{{\rm AL},k}^{\rm H}
  \big(\sigma_{\bm{n}_k}^2 \bm{I} + P_k \bm{\Psi}_k\big) \bm{F}_{{\rm AL},k}\Big)^{\frac{1}{2}}
  \bm{F}_{{\rm D},k} \big(\bar{\bm{F}}_{{\rm AR},k} \bar{\bm{F}}_{{\rm AR},k}^{\rm H}\big)^{\frac{1}{2}}
  \widetilde{\bm{U}}_k^{\rm H} ,\label{eq59} \\
 \bm{\Pi}_{{\rm R},k} =& \big(\bar{\bm{F}}_{{\rm AR},k} \bar{\bm{F}}_{{\rm AR},k}^{\rm H}\big)^{-\frac{1}{2}}
  \bar{\bm{F}}_{{\rm AR},k} , \label{eq60} \\
 \bm{\Pi}_{{\rm L},k} =& \big(\sigma_{\bm{n}_k}^2\bm{I}+P_k\bm{\Psi}_k\big)^{\frac{1}{2}}\bm{F}_{{\rm AL},k}
  \Big( \bm{F}_{{\rm AL},k}^{\rm H} \big(\sigma_{\bm{n}_k}^2\bm{I}+P_k\bm{\Psi}_k\big)
  \bm{F}_{{\rm AL},k}\Big)^{-\frac{1}{2}} ,\label{eq61}
\end{align}
 where $\widetilde{\bm{U}}_k$ is a left unitary matrix of appropriate dimension yet to
 be determined, the optimization problem (\ref{E_Opt_Medium_A}) can be reformulated into
\begin{align}\label{E_Opt_Medium_B} 
\!\!\! \begin{array}{cl}
 \max\limits_{\bar{\bm{F}}_k} \!\!\! &\!\! \bm{\lambda}\Big\{\! \bm{\Pi}_{{\rm R},k}^{\rm H} \widetilde{\bm{U}}_k^{\rm H}
 \widetilde{\bm{F}}_{{\rm D},k}^{\rm H} \bm{\Pi}_{{\rm L},k}^{\rm H} \big(\sigma_{\bm{n}_k}^2
 \bm{I}\! +\! P_k \bm{\Psi}_k\big)^{-\frac{1}{2}} \widehat{\bm{H}}_k^{\rm H} \widehat{\bm{H}}_k 
 \big(\sigma_{\bm{n}_k}^2 \bm{I}\! +\! P_k \bm{\Psi}_k\big)^{-\frac{1}{2}}
 \bm{\Pi}_{{\rm L},k} \widetilde{\bm{F}}_{{\rm D},k} \widetilde{\bm{U}}_k \bm{\Pi}_{{\rm R},k}\Big\} ,\! \\
 {\rm{s.t.}}\!\!\! &\!\!  {\rm{Tr}}\big(\widetilde{\bm{F}}_{{\rm D},k}\widetilde{\bm{F}}_{{\rm D},k}^{\rm H} \big)\le P_k ,
  \ \bm{F}_{{\rm AL},k}\in \mathcal{F}_{{\rm PL},k}, \ \bm{F}_{{\rm AR},k}\in \mathcal{F}_{{\rm PR},k}.
\end{array}\!\!
\end{align}

 Similar to point-to-point MIMO systems \cite{matrixMonoOpt2018}, we have the following
 two conclusions.

\begin{conclusion}\label{C4}
 Based on the definition of $\bm{\Pi}_{{\rm R},k}$ in (\ref{eq60}), it can be concluded that the
 singular values of $\bar{\bm{F}}_{{\rm AR},k}$ do not affect the system performance. The
 right singular vectors of the optimal $\bar{\bm{F}}_{{\rm AR},k}$ correspond to the left
 singular vectors of the preceding-hop channel, i.e., $ \bm{U}_{\bm{\mathcal{H}}_k}^{\rm H} $.
\end{conclusion}

\begin{conclusion}\label{C5}
 Based on the SVDs
\begin{align}
 & \!\!\!\!\! \widehat{\bm{H}}_k\big(\sigma_{\bm{n}_k}^2\! \bm{I}\! +\! P_k
  \bm{\Psi}_k)^{-\frac{1}{2}} \bm{\Pi}_{{\rm L},k}
  \widetilde{\bm{F}}_{{\rm D},k}\! =\! \widetilde{\bm{U}}_k \widetilde{\bm{\Lambda}}_k \widetilde{\bm{V}}_k^{\rm H}
  \ \text{with} \ \widetilde{\bm{\Lambda}}_k\!\! \searrow , \label{eq63} \\
 & \bm{\Pi}_{{\rm R},k} = \bm{U}_{\bm{\Pi}_{{\rm R},k}} \bm{\Lambda}_{\bm{\Pi}_{{\rm R},k}}
  \bm{V}_{\bm{\Pi}_{{\rm R},k}}^{\rm H} \ \text{with} \ \bm{\Lambda}_{\bm{\Pi}_{{\rm R},k}}\!\! \searrow , \label{eq64}
\end{align}
 the optimal ${\widetilde{\bm{U}}}_k$ equals to
\begin{align}\label{eq65}
 {\widetilde{\bm{U}}}_{k,{\rm{opt}}} =& \widetilde{\bm{V}}_k \bm{U}_{\bm{\Pi}_{{\rm R},k}}^{\rm H} .
\end{align}
\end{conclusion}

 Based on Conclusions~\ref{C4} and \ref{C5}, the optimization problem (\ref{E_Opt_Medium_B})
 is equivalent to the much simpler one as follows
\begin{align}\label{eq66}
 \begin{array}{cl}
 \max\limits_{\bar{\bm{F}}_k} \!\! & \bm{\lambda}\Big\{ \widetilde{\bm{F}}_{{\rm D},k}^{\rm H}
  \bm{\Pi}_{{\rm L},k}^{\rm H} \big(\sigma_{\bm{n}_k}^2 \bm{I} + P_k \bm{\Psi}_k\big)^{-\frac{1}{2}}
  \widehat{\bm{H}}_k^{\rm H} \widehat{\bm{H}}_k  \big(\sigma_{\bm{n}_k}^2 \bm{I} + P_k \bm{\Psi}_k\big)^{-\frac{1}{2}}
   \bm{\Pi}_{{\rm L},k} \widetilde{\bm{F}}_{{\rm D},k}\Big\} , \\
 {\rm{s.t.}} \!\! &  {\rm{Tr}}\big(\widetilde{\bm{F}}_{{\rm D},k}\widetilde{\bm{F}}_{{\rm D},k}^{\rm H} \big)\le P_k , 
  \ \bm{F}_{{\rm AL},k}\in \mathcal{F}_{{\rm PL},k}, \ \bm{F}_{{\rm AR},k}\in \mathcal{F}_{{\rm PR},k} ,
\end{array}
\end{align}
 which further equals to the following matrix monotonic optimization problem
\begin{align}\label{MM_medium_hop} 
 \begin{array}{cl}
 \max\limits_{\bar{\bm{F}}_k} \!\! &  \widetilde{\bm{F}}_{{\rm D},k}^{\rm H}
  \bm{\Pi}_{{\rm L},k}^{\rm H} \big(\sigma_{\bm{n}_k}^2 \bm{I} + P_k \bm{\Psi}_k\big)^{-\frac{1}{2}}
  \widehat{\bm{H}}_k^{\rm H} \widehat{\bm{H}}_k  \big(\sigma_{\bm{n}_k}^2 \bm{I} + P_k \bm{\Psi}_k\big)^{-\frac{1}{2}}
   \bm{\Pi}_{{\rm L},k} \widetilde{\bm{F}}_{{\rm D},k} , \\
 {\rm{s.t.}} \ &  {\rm{Tr}}\big(\widetilde{\bm{F}}_{{\rm D},k}\widetilde{\bm{F}}_{{\rm D},k}^{\rm H} \big)\le P_k , 
  \  \bm{F}_{{\rm AL},k}\in \mathcal{F}_{{\rm PL},k}, \ \bm{F}_{{\rm AR},k}\in \mathcal{F}_{{\rm PR},k} .
\end{array}
\end{align}
 Similar to the matrix monotonic optimization (\ref{MM_first_hop}), we can obtain
 the optimal solution of (\ref{MM_medium_hop}).

\begin{conclusion}\label{C6}
 The singular values of $\big(\sigma_{\bm{n}_k}^2\bm{I}+P_k\bm{\Psi}_k\big)^{\frac{1}{2}}
 \bm{F}_{{\rm AL},k}$ do not affect the system performance. The left eigenvectors of the SVD
 for $\big(\sigma_{\bm{n}_k}^2\bm{I}+P_k\bm{\Psi}_1\big)^{\frac{1}{2}}\bm{F}_{{\rm AL},k}$
 have the maximum inner product with respect to the eigenvectors $\bm{V}_{\bm{\mathcal{H}}_k}$
 defined by the following SVD
\begin{align}\label{eq68}
 \widehat{\bm{H}}_k\big(\sigma_{\bm{n}_k}^2\bm{I}\! +\! P_k\bm{\Psi}_k\big)^{-\frac{1}{2}}\! =\!
  \bm{U}_{\bm{\mathcal{H}}_k}\bm{\Lambda}_{\bm{\mathcal{H}}_k}\bm{V}_{\bm{\mathcal{H}}_k}^{\rm H}
  \ \text{with} \ \bm{\Lambda}_{\bm{\mathcal{H}}_k}\! \searrow .\!
\end{align}
\end{conclusion}

\begin{conclusion}\label{C7}
 Based on the SVD
\begin{align}\label{eq69}
 \widehat{\bm{H}}_k\big(\sigma_{\bm{n}_k}^2\bm{I}\! +\! P_k\bm{\Psi}_k\big)^{-\frac{1}{2}}
  \bm{\Pi}_{{\rm L},k}\! =\! \bm{U}_{\bm{\Pi},k} \bm{\Lambda}_{\bm{\Pi},k} \bm{V}_{\bm{\Pi},k}^{\rm H}
  \, \text{with} \, \bm{\Lambda}_{\bm{\Pi},k} \! \searrow ,\!
\end{align}
 for given $\bm{F}_{{\rm AL},k}$, all the Pareto optimal $\widetilde{\bm{F}}_{{\rm D},k}$ of
 the optimization problem (\ref{MM_medium_hop}) satisfy the structure:
\begin{align}\label{eq70}
 \widetilde{\bm{F}}_{{\rm D},k} =& \bm{V}_{\bm{\Pi},k}\bm{\Lambda}_{\widetilde{\bm{F}}_{{\rm D},k}}
  \bm{U}_{\rm Arb}^{\rm H} ,
\end{align}
 where $\bm{\Lambda}_{\widetilde{\bm{F}}_{{\rm D},k}}$ is a rectangular diagonal matrix.
\end{conclusion}	

 Based on Conclusion~\ref{C7} and the definition in (\ref{eq59}), after computing the optimal
 $\widetilde{\bm{F}}_{{\rm D},k}$, the optimal $\bm{F}_{{\rm D},k}$ is given by
\begin{align}\label{eq71}
 \bm{F}_{{\rm D},k} =& \sqrt{\frac{P_k}{\alpha_k}} \Big( \bm{F}_{{\rm AL},k}^{\rm H}
  \big(\sigma_{\bm{n}_k}^2 \bm{I} + P_k \bm{\Psi}_k\big) \bm{F}_{{\rm AL},k}\Big)^{-\frac{1}{2}}
  \widetilde{\bm{F}}_{{\rm D},k} \widetilde{\bm U}_k \big(\bar{\bm{F}}_{{\rm AR},k}
  \bar{\bm{F}}_{{\rm AR},k}^{\rm H}\big)^{-\frac{1}{2}} ,
\end{align}
 where
\begin{align}\label{eq72}
 \alpha_k\! =& {\rm{Tr}}\bigg(\! \Big( \bm{F}_{{\rm AL},k}^{\rm H} \big( \sigma_{\bm{n}_k}^2\bm{I}
  \! +\! P_k\bm{\Psi}_k\big) \bm{F}_{{\rm AL},k}\Big)^{-\frac{1}{2}} \! \bm{F}_{{\rm AL},k}^{\rm H}
  \bm{F}_{{\rm AL},k}  \Big(\! \bm{F}_{{\rm AL},k}^{\rm H} \big( \sigma_{\bm{n}_k}^2\! \bm{I}\!
  +\! P_k\bm{\Psi}_k\big) \bm{F}_{{\rm AL},k}\Big)^{-\frac{1}{2}} \widetilde{\bm{F}}_{{\rm D},k}
  \widetilde{\bm{F}}_{{\rm D},k}^{\rm H}\! \bigg)\! .\!
\end{align}
	
\subsection{Final Hop}\label{S5.3}

 By noting the definition in (\ref{eq53}), the optimization problem
 (\ref{Vector_Opt}) for the final $K$th hop becomes
\begin{align}\label{E_Opt_Final} 
 \begin{array}{cl}
 \max\limits_{\bm{G}_{\rm A},\bar{\bm{F}}_K} \!\! &  \bm{\lambda}\Big\{\bar{\bm{F}}_{{\rm AR},K}^{\rm H}
  \bm{F}_{{\rm D},K}^{\rm H} \bm{F}_{{\rm AL},K}^{\rm H} \widehat{\bm{H}}_K^{\rm H} \bm{G}_{\rm A}^{\rm H}
  \bm{K}_{\bm{n}_K}^{-1}\bm{G}_{\rm A} \widehat{\bm{H}}_K \bm{F}_{{\rm AL},K} \bm{F}_{{\rm D},K}
  \bar{\bm{F}}_{{\rm AR},K} \Big\} , \\
 {\rm{s.t.}} \!\! & {\rm{Tr}}\big( \bm{F}_{{\rm AL},K} \bm{F}_{{\rm D},K} \bar{\bm{F}}_{{\rm AR},K}
  \bar{\bm{F}}_{{\rm AR},K}^{\rm H} \bm{F}_{{\rm D},K}^{\rm H} \bm{F}_{{\rm AL},K}^{\rm H}\big)\! \le\! P_K , \\
 \!\! & \bm{F}_{{\rm AL},K}\in \mathcal{F}_{{\rm PL},K}, \, \bm{F}_{{\rm AR},K}\in \mathcal{F}_{{\rm PR},K} , \,
  \bm{G}_{\rm A} \in \mathcal{F}_{\rm G} ,
\end{array}
\end{align}
 where the equivalent noise covariance matrix is given by
\begin{align}\label{eq74}
 \bm{K}_{\bm{n}_K}\! =& \bm{G}_{\rm A} \Big( \Big( \sigma_{\bm{n}_K}^2\! +\!
  {\rm Tr} \big( \bm{F}_{{\rm AL},K} \bm{F}_{{\rm D},K} \bar{\bm{F}}_{{\rm AR},K}
  \bar{\bm{F}}_{{\rm AR},K}^{\rm H} \bm{F}_{{\rm D},K}^{\rm H} \bm{F}_{{\rm AL},K}^{\rm H}
  \bm{\Psi}_K\big) \Big) \bm{I} \Big) \bm{G}_{\rm A}^{\rm H} \triangleq
  \eta_K \bm{G}_{\rm A} \bm{G}_{\rm A}^{\rm H} .
\end{align}

 Clearly, the power constraint in (\ref{E_Opt_Final}), namely,
\begin{align}\label{eq75}
 {\rm{Tr}}\big( \bm{F}_{{\rm AL},K} \bm{F}_{{\rm D},K} \bar{\bm{F}}_{{\rm AR},K}
  \bar{\bm{F}}_{{\rm AR},K}^{\rm H} \bm{F}_{{\rm D},K}^{\rm H} \bm{F}_{{\rm AL},K}^{\rm H}\big) \le P_K ,
\end{align}
 is equivalent to
\begin{align}\label{eq76}
 & \frac{{\rm Tr}\Big(\! \big(\sigma_{\bm{n}_K}^2\bm{I}\! +\! P_K\bm{\Psi}_K\big)\bm{F}_{{\rm AL},K}
  \bm{F}_{{\rm D},K}\bar{\bm{F}}_{{\rm AR},K} \bar{\bm{F}}_{{\rm AR},K}^{\rm H}\bm{F}_{{\rm D},K}^{\rm H}
  \bm{F}_{{\rm AL},K}^{\rm H}\! \Big)}{\eta_K} \le P_K .
\end{align}
 By defining the following auxiliary variables
\begin{align}
 \widetilde{\bm{F}}_{{\rm D},K} =& \eta_K^{-\frac{1}{2}}\Big(\bm{F}_{{\rm AL},K}^{\rm H}
  \big(\sigma_{\bm{n}_K}^2 \bm{I} + P_K\bm{\Psi}_K\big) \bm{F}_{{\rm AL},K}\Big)^{\frac{1}{2}}
  \bm{F}_{{\rm D},K}  \Big(\bar{\bm{F}}_{{\rm AR},K} \bar{\bm{F}}_{{\rm AR},K}^{\rm H}\Big)^{\frac{1}{2}}
  \widetilde{\bm{U}}_K^{\rm H}, \label{eq77} \\
 \bm{\Pi}_{{\rm R},K} =& \Big( \bar{\bm{F}}_{{\rm AR},K} \bar{\bm{F}}_{{\rm AR},K}^{\rm H}\Big)^{-\frac{1}{2}}
  \bar{\bm{F}}_{{\rm AR},K} , \label{eq78} \\
 \bm{\Pi}_{{\rm L},K} =& \big(\sigma_{\bm{n}_K}^2\bm{I}+P_K\bm{\Psi}_K\big)^{\frac{1}{2}}
  \bm{F}_{{\rm AL},K}\Big( \bm{F}_{{\rm AL},K}^{\rm H}  \big(\sigma_{\bm{n}_K}^2\bm{I}
  + P_K\bm{\Psi}_K\big) \bm{F}_{{\rm AL},K}\Big)^{-\frac{1}{2}} , \label{eq79} \\
 \bm{\Pi}_{\rm G} =& \big(\bm{G}_{\rm A}\bm{G}_{\rm A}^{\rm H}\big)^{-\frac{1}{2}} \bm{G}_{\rm A} , \label{eq80}
\end{align}
 where $\widetilde{\bm{U}}_K$ is a left unitary matrix yet to be determined, the vector
 optimization problem (\ref{E_Opt_Final}) can be reformulated as
\begin{align}\label{E_Opt_Final_A} 
 \begin{array}{cl}
 \max\limits_{\bm{G}_{\rm A},\bar{\bm{F}}_K} \!\! & \bm{\lambda}\Big\{\!
  \bm{\Pi}_{{\rm R},K}^{\rm H}\widetilde{\bm{U}}_K^{\rm H}\widetilde{\bm{F}}_{{\rm D},K}^{\rm H}
  \bm{\Pi}_{{\rm L},K}^{\rm H}\big(\sigma_{\bm{n}_K}^2\bm{I}\! +\! P_K\bm{\Psi}_K\! \big)^{-\frac{1}{2}}
  \widehat{\bm{H}}_K^{\rm H}\bm{\Pi}_{\rm G}^{\rm H} \\
 \!\! & \times \bm{\Pi}_{\rm G}\widehat{\bm{H}}_K\big(\sigma_{\bm{n}_K}^2\bm{I}\! +\!
   P_K\bm{\Psi}_K\! \big)^{-\frac{1}{2}} \bm{\Pi}_{{\rm L},K}\widetilde{\bm{F}}_{{\rm D},K}
  \widetilde{\bm{U}}_K\bm{\Pi}_{{\rm R},K}\! \Big\} , \\
 {\rm{s.t.}} \!\! & {\rm{Tr}}\big( \widetilde{\bm{F}}_{{\rm D},K}
  \widetilde{\bm{F}}_{{\rm D},K}^{\rm H}\big)\! \le\! P_K, \ \bm{F}_{{\rm AL},K}\in \mathcal{F}_{{\rm PL},K},
  \ \bm{F}_{{\rm AR},K}\in \mathcal{F}_{{\rm PR},K} , \,
  \bm{G}_{\rm A} \in \mathcal{F}_{\rm G}.\!
\end{array}
\end{align}
 Then we readily  have the following two conclusions.

\begin{conclusion}\label{C8}
 Based on the definition of $\bm{\Pi}_{{\rm R},K}$, it can be concluded that the
 singular values of $\bar{\bm{F}}_{{\rm AR},K} $ do not affect the system performance.
 The right singular vectors of the optimal $\bar{\bm{F}}_{{\rm AR},K}$ correspond to
 the left singular vectors of the preceding-hop channel, i.e.,
 $\bm{U}_{\bm{\mathcal{H}}_K}^{\rm H}$.
\end{conclusion}

\begin{conclusion}\label{C9}
 Based on the SVDs
\begin{align}
 & \widehat{\bm{H}}_K\big(\sigma_{\bm{n}_K}^2\! \bm{I}\! +\! P_K\bm{\Psi}_K\big)^{-\frac{1}{2}}
  \bm{\Pi}_{{\rm L},K}\widetilde{\bm{F}}_{{\rm D},K}\!\! =\! \widetilde{\bm{U}}_K
  \widetilde{\bm{\Lambda}}_K \widetilde{\bm{V}}_K^{\rm H} \,
  \text{with} \, \widetilde{\bm{\Lambda}}_K\!\! \searrow  ,\! \label{eq82} \\
 & \bm{\Pi}_{{\rm R},K} = \bm{U}_{\bm{\Pi}_{{\rm R},K}} \bm{\Lambda}_{\bm{\Pi}_{{\rm R},K}}
  \bm{V}_{\bm{\Pi}_{{\rm R},K}}^{\rm H} \ \text{with} \ \bm{\Lambda}_{\bm{\Pi}_{{\rm R},K}} \searrow ,
  \label{eq83}
\end{align}
 the optimal $\widetilde{\bm{U}}_K$ is derived as
\begin{align}\label{eq84}
 \widetilde{\bm{U}}_{K,{\rm opt}} =& \widetilde{\bm{V}}_K \bm{U}_{\bm{\Pi}_{{\rm R},K}}^{\rm H} .
\end{align}
\end{conclusion}
	
 Based on Conclusions~\ref{C8} and \ref{C9}, the optimization (\ref{E_Opt_Final_A}) can
 be simplified into:	
\begin{align}\label{eq85}
 \begin{array}{cl}
 \max\limits_{\bm{G}_{\rm A},\bar{\bm{F}}_K}\!\! &\!\! \bm{\lambda}\Big\{\!
  \widetilde{\bm{F}}_{{\rm D},K}^{\rm H} \bm{\Pi}_{{\rm L},K}^{\rm H} \big(\sigma_{\bm{n}_K}^2
  \bm{I}\! +\! P_K \bm{\Psi}_K\big)^{-\frac{1}{2}} \widehat{\bm{H}}_K^{\rm{H}} \bm{\Pi}_{\rm G}^{\rm H}
  \bm{\Pi}_{\rm G} \widehat{\bm{H}}_K\big(\sigma_{\bm{n}_K}^2 \bm{I}\! +\! P_K
  \bm{\Psi}_K)^{-\frac{1}{2}} \bm{\Pi}_{{\rm L},K} \widetilde{\bm{F}}_{{\rm D},K}\! \Big\}\! , \! \\
 {\rm{s.t.}}\!\! &\!\! {\rm{Tr}}\big( \widetilde{\bm{F}}_{{\rm D},K}
  \widetilde{\bm{F}}_{{\rm D},K}^{\rm H}\big)\! \le\! P_K\! , \  \bm{F}_{{\rm AL},K}\in \mathcal{F}_{{\rm PL},K},
  \, \bm{F}_{{\rm AR},K}\in \mathcal{F}_{{\rm PR},K} , \,
  \bm{G}_{\rm A} \in \mathcal{F}_{\rm G} .\!
\end{array}\!\!
\end{align}
 The optimization (\ref{eq85}) is equivalent to the following matrix monotonic optimization
 problem
\begin{align}\label{MM_final_hop} 
 \begin{array}{cl}
 \max\limits_{\bm{G}_{\rm A},\bar{\bm{F}}_K}\!\! &
  \widetilde{\bm{F}}_{{\rm D},K}^{\rm H} \bm{\Pi}_{{\rm L},K}^{\rm H} \big(\sigma_{\bm{n}_K}^2
  \bm{I}\! +\! P_K \bm{\Psi}_K\big)^{-\frac{1}{2}} \widehat{\bm{H}}_K^{\rm{H}} \bm{\Pi}_{\rm G}^{\rm H}
  \bm{\Pi}_{\rm G} \widehat{\bm{H}}_K\big(\sigma_{\bm{n}_K}^2 \bm{I}\! +\! P_K
  \bm{\Psi}_K)^{-\frac{1}{2}} \bm{\Pi}_{{\rm L},K} \widetilde{\bm{F}}_{{\rm D},K} ,\! \\
 {\rm{s.t.}}\!\! & {\rm{Tr}}\big( \widetilde{\bm{F}}_{{\rm D},K}
  \widetilde{\bm{F}}_{{\rm D},K}^{\rm H}\big)\! \le\! P_K\! , \ \bm{F}_{{\rm AL},K}\in \mathcal{F}_{{\rm PL},K},
  \, \bm{F}_{{\rm AR},K}\in \mathcal{F}_{{\rm PR},K} , \,
  \bm{G}_{\rm A} \in \mathcal{F}_{\rm G} , \!
\end{array}\!\!
\end{align}
 and we readily have the following three conclusions.

\begin{conclusion}\label{C10}
 The singular values of the matrix $\big(\sigma_{\bm{n}_K}^2\bm{I} + P_K\bm{\Psi}_K \big)^{\frac{1}{2}}
 \bm{F}_{{\rm AL},K}$ do not affect the system performance. The left eigenvectors of the SVD for
 $\big(\sigma_{\bm{n}_K}^2\bm{I}+P_K\bm{\Psi}_K\big)^{\frac{1}{2}}\bm{F}_{{\rm AL},K}$
 have the maximum inner product with respect to the eigenvectors $\bm{V}_{\bm{\mathcal{H}}_K}$
 defined by the following SVD
\begin{align}\label{eq87}
 \widehat{\bm{H}}_K\big(\sigma_{\bm{n}_K}^2\bm{I}\! +\! P_K\bm{\Psi}_K)^{-\frac{1}{2}}\! =\!
  \bm{U}_{\bm{\mathcal{H}}_K} \bm{\Lambda}_{\bm{\mathcal{H}}_K} \bm{V}_{\bm{\mathcal{H}}_K}^{\rm H}
  \, \text{with} \, \bm{\Lambda}_{\bm{\mathcal{H}}_K}\! \searrow .
\end{align}
\end{conclusion}

\begin{conclusion}\label{C11}
 The singular values of $\bm{G}_{\rm A}$ do not affect the system performance. The right
 eigenvectors of the SVD for $\bm{G}_{\rm A}$ have the maximum inner product w.r.t.
 the eigenvectors $\bm{U}_{\bm{\mathcal{H}}_K}$.
\end{conclusion}

\begin{conclusion}\label{C12}
 Based on the SVD
\begin{align}\label{eq88}
 \bm{\Pi}_{\rm G}\widehat{\bm{H}}_K\big(\sigma_{\bm{n}_K}^2\bm{I}\! +\! P_K\bm{\Psi}_K\big)^{-\frac{1}{2}}
  \bm{\Pi}_{{\rm L},K}\! =& \bm{U}_{\bm{\Pi},K} \bm{\Lambda}_{\bm{\Pi},K} \bm{V}_{\bm{\Pi},K}^{\rm H}
\ \text{with} \ \bm{\Lambda}_{\bm{\Pi},K}\! \searrow ,\!
\end{align}
 for given $\bm{F}_{{\rm AL},K}$, all the Pareto optimal solutions $\widetilde{\bm{F}}_{{\rm D},K}$
 of the optimization problem (\ref{MM_final_hop}) satisfy the following structure
\begin{align}\label{eq89}
 \widetilde{\bm{F}}_{{\rm D},K} = \bm{V}_{\bm{\Pi},K} \bm{\Lambda}_{\widetilde{\bm{F}}_{{\rm D},K}}
  \bm{U}_{\rm Arb}^{\rm H} ,
\end{align}
 where $\bm{\Lambda}_{\widetilde{\bm{F}}_{{\rm D},K}}$ is a rectangular diagonal matrix
 and $\bm{U}_{\rm Arb}$ is an arbitrary unitary matrix with a proper dimension.
\end{conclusion}

 Based on the definition in (\ref{eq77}), when the optimal $\widetilde{\bm{F}}_{{\rm D},K}$
 is given, the optimal $\bm{F}_{{\rm D},K}$ can be computed according to
\begin{align}\label{eq90}
 \bm{F}_{{\rm D},K} =& \sqrt{\frac{P_K}{\alpha_K}}\Big( \bm{F}_{{\rm AL},K}^{\rm H}
  \big(\sigma_{\bm{n}_K}^2\bm{I}+P_K\bm{\Psi}_K\big)\bm{F}_{{\rm AL},K}\Big)^{-\frac{1}{2}}
  \widetilde{\bm{F}}_{{\rm D},K} \widetilde{\bm{U}}_K \big(\bar{\bm{F}}_{{\rm AR},K}
  \bar{\bm{F}}_{{\rm AR},K}^{\rm H})^{-\frac{1}{2}} ,
\end{align}
 where $\alpha_K$ is given by
\begin{align}\label{eq91}
 \alpha_K =& {\rm{Tr}}\Big(\Big(\bm{F}_{{\rm AL},K}^{\rm H}\big(\sigma_{\bm{n}_K}^2\bm{I}\! +\!
  P_K\bm{\Psi}_K\big)\bm{F}_{{\rm AL},K}\Big)^{-\frac{1}{2}}\bm{F}_{{\rm AL},K}^{\rm H}
   \bm{F}_{{\rm AL},K} \nonumber \\
  &  \times \Big(\!\bm{F}_{{\rm AL},K}^{\rm H}\big(\sigma_{\bm{n}_K}^2\bm{I}\! +\!
   P_K\bm{\Psi}_K\big)\bm{F}_{{\rm AL},K}\! \Big)^{-\frac{1}{2}} \widetilde{\bm{F}}_{{\rm D},K}^{\rm H}
   \widetilde{\bm{F}}_{{\rm D},K}\!\Big)\! .\!
\end{align}

\section{Analog Transceiver Optimizations}\label{S6}

 In the previous section, the optimal structures of the hybrid transceivers are
 derived. Due to the physical limitations of analog transceivers, the processing
 factors corresponding to individual analog antenna elements are constrained to
 be unit-modulus. Thus, it is the primary concern to derive an efficient
 algorithm to design analog transceivers based on the obtained optimal structures.
 In addition, for multi-hop AF MIMO relaying systems, the low complex analog
 beamforming is of practical interest as well. This section focuses on these
 critical issues.

\subsection{Proposed Analog Beamformer Design}\label{S6.1}

\subsubsection{Analog Transmit Precoder Design}

 Start from the analog precoder design. In Section~\ref{S5}, it is shown that
 the auxiliary variable of the beamformer at the $k$th relay takes
 the following form
\begin{align}\label{eq-uniform} 
 \bm{\Pi}_{{\rm L},k} \! =& \bm{D}_k \bm{F}_{\mathrm{AL},k}
  \left( \bm{F}_{\mathrm{AL},k}^{\rm H} \bm{D}_k^{\rm H} \bm{D}_k
  \bm{F}_{\mathrm{AL},k} \right)^{-\frac{1}{2}} , ~ 1\le k\le K ,\!
\end{align}
 where $\bm{D}_k$ is any invertible matrix with appropriate dimension, and
 $\bm{F}_{\mathrm{AL},k}$ is the analog transmit precoder to be designed. It is
 interesting to point out that (\ref{eq-uniform}) is actually a general form of
 analog beamformer design problem, and thus can also be utilized to other
 beamformer design situations. In Section~\ref{S5}, it has been proven that only the
 left singular matrix of $\bm{\Pi}_{{\rm L},k}$, which is also equivalent to the
 left singular matrix of $\bm{D}_k\bm{F}_{\mathrm{AL},k}$, should correspond with
 the eigenvectors of the channel. More specifically, given the following SVD
\begin{align}\label{eq93}
  \bm{D}_{k}\bm{F}_{\mathrm{AL},k} =& \bm{U}_{{\rm L},k} \bm{\Sigma}_{\mathrm{L},k}
  \bm{V}_{\mathrm{L},k}^{\rm H},
\end{align}
 the optimal solution of $\bm{U}_{\mathrm{L},k}$ is $ \bm{V}_{\bm{\mathcal{H}}_k} $.
 
 Instinctively, the angle matrix of the desired value of $\bm{\Pi}_{\mathrm{L},k}$,
 namely, $\bm{V}_{\bm{\mathcal{H}}_k}$, could be used to compose the analog
 beamformer, which is denoted as $\mathcal{P}_{\mathcal{F}}\big(\bm{D}_{k}^{-1}
 \bm{V}_{\bm{\mathcal{H}}_k}\big)$. However, as the amplitude information is missing
 by this method, the performance of transceivers designed by such rudimentary idea
 could be poor. An improved design is to minimize the Frobenius norm of the error
 between the desired full digital solution and the unit-modulus beamformer.
 
 Different from the previous work \cite{matrixMonoOpt2018}, we consider a more general
 design form to further improve the performance. Note that the system
 performance can be guaranteed if the first $N$ channel eigenvectors are perfectly
 matched. However, due to the limitation of hybrid structure, there is always a performance
 gap between the optimal full digital transceivers and the hybrid transceivers. On the
 other hand, in practice, the number of RF chains is often larger than that of data
 streams. A reasonable instinct is to utilize extra design freedom offered by
 these extra RF chains to enhance the matching accuracy. Moreover, in this paper, a
 weighted norm is utilized to account for the varying influence of different
 bases in the signal space. In this way, the associated optimization problem can be
 formulated as
\begin{align}\label{prob-analog-org} 
\!\!\!\!\begin{array}{cl}
  \min\limits_{\bm{\Sigma}_{\mathrm{L},k},\bm{V}_{\mathrm{L},k},\bm{F}_{\mathrm{AL},k}}\!\! &\!\!
   \Big\lVert \bm{W}_k^{\frac{1}{2}} \big( \bm{V}_{\bm{\mathcal{H}}_k} \bm{\Sigma}_{\mathrm{L},k}
   \bm{V}_{\mathrm{L},k}^{\rm H}\! -\! \bm{D}_k \bm{F}_{\mathrm{AL},k} \big) \Big\rVert_F^2 , \\
  \text{s.t.}\!\! &\!\! \bm{V}_{\mathrm{L},k}\in \mathcal{U} , \bm{F}_{\mathrm{AL},k} \in
    \mathcal{F}_{\mathrm{PL},k} , \
    \bm{\Sigma}_{\mathrm{L},k} = \mathrm{diag} \{ \sigma_{\mathrm{L},1}, \cdots , \sigma_{\mathrm{L},K} \} ,
\end{array}\!
\end{align}
 where $\mathcal{U}\! =\! \{\bm{U} | \bm{U}\bm{U}^{\rm H}\! =\! \bm{U}^{\rm H}\bm{U}\!
 =\! \bm{I}\}$ is the unitary matrix set. We can choose the weight matrix as
 $\bm{W}_k\! =\! \bm{V}_{\bm{\mathcal{H}}_k}\bm{\Lambda}_{\bm{W}_k}
 \bm{V}_{\bm{\mathcal{H}}_k}^{\rm H}$ in which $\bm{\Lambda}_{\bm{W}_k}$ is a diagonal
 matrix. Then, denote
\begin{align}
 \bm{V}_{\bm{\mathcal{H}}_k} =& \begin{bmatrix}
   \widetilde{\bm{V}}_{\bm{\mathcal{H}}_k} & \widehat{\bm{V}}_{\bm{\mathcal{H}}_k}
   \end{bmatrix} , \label{eq95} \\
 \bm{\Sigma}_{\mathrm{L},k} =& \begin{bmatrix}
  \widetilde{\bm{\Sigma}}_{\mathrm{L},k} & \\
  & \widehat{\bm{\Sigma}}_{\mathrm{L},k}
  \end{bmatrix} , \label{eq96} \\
 \bm{V}_{\mathrm{L},k} =& \begin{bmatrix}
  \widetilde{\bm{V}}_{\mathrm{L},k}, & \widehat{\bm{V}}_{\mathrm{L},k}
  \end{bmatrix} , \label{eq97}
\end{align}
 where $\widetilde{\bm{\Sigma}}_{\mathrm{L},k}\! \in\! \mathbb{R}^{N\times N}$ and
 $\widehat{\bm{\Sigma}}_{\mathrm{L},k}\! \in\! \mathbb{C}^{(N-N_{\rm RF})\times (N-N_{\rm RF})}$
 are diagonal matrices, while $\widetilde{\bm{V}}_{\bm{\mathcal{H}}_k}\! \in\!
 \mathbb{C}^{N_{t,k}\times N}$, $\widehat{\bm{V}}_{\bm{\mathcal{H}}_k}\! \in\!
 \mathbb{C}^{N_{t,k}\times (N_{\rm RF}-N)}$, $\widetilde{\bm{V}}_{\mathrm{L},k}
 \! \in\! \mathbb{C}^{N_{\rm RF}\times N}$ and $\widehat{\bm{V}}_{\mathrm{L},k}
 \! \in\! \mathbb{C}^{N_{\rm RF}\times (N_{\rm RF}-N)}$ are complex matrices.
 Thus the objective function of (\ref{prob-analog-org}) can be transformed into
\begin{align}\label{eq98}
 \Big\lVert \bm{W}_k^{\frac{1}{2}} \big( \bm{V}_{\bm{\mathcal{H}}_k} \bm{\Sigma}_{\mathrm{L},k}
  \bm{V}_{\mathrm{L},k}^{\rm H}\! -\! \bm{D}_k \bm{F}_{\mathrm{AL},k} \big) \Big\rVert_F^2
  \! = \! \Big\lVert \bm{W}_k^{\frac{1}{2}} \big( \widetilde{\bm{V}}_{\bm{\mathcal{H}}_k}
  \widetilde{\bm{\Sigma}}_{\mathrm{L},k} \widetilde{\bm{V}}_{\mathrm{L},k}^{\rm H}\! +\!
  \widehat{\bm{V}}_{\bm{\mathcal{H}}_k} \widehat{\bm{\Sigma}}_{\mathrm{L},k}
  \widehat{\bm{V}}_{\mathrm{L},k}^{\rm H}\! -\! \bm{D}_k \bm{F}_{\mathrm{AL},k} \big) \Big\rVert_F^2
  \! . \!
\end{align}
 
 It is worth highlighting that there is no constraint imposed on the matrix variable
 $\bm{\Sigma}_{\mathrm{L},k}$ in the optimization (\ref{prob-analog-org}). As a result,
 the optimal $\widetilde{\bm{\Sigma}}_{\mathrm{L},k}$ and $\widehat{\bm{\Sigma}}_{\mathrm{L},k}$
 can be derived in closed-form as
\begin{align} 
 \widetilde{\bm{\Sigma}}_{\mathrm{L},k} =& \Big( \mathrm{diag} \big\{
  \bm{d}\big[\widetilde{\bm{V}}_{\bm{\mathcal{H}}_k}^{\rm H} \bm{W}_k \widetilde{\bm{V}}_{\bm{\mathcal{H}}_k} \big]
  \big\}\Big)^{-1} \Re \Big\{ \mathrm{diag} \big\{ \bm{d}\big[ \widetilde{\bm{V}}_{\bm{\mathcal{H}}_k}^{\rm H}
  \bm{W}_k \bm{D}_k \bm{F}_{\mathrm{AL},k }  \widetilde{\bm{V}}_{\mathrm{L},k}
  \big]\big\} \Big\} , \label{eq-diag-left} \\
 \widehat{\bm{\Sigma}}_{\mathrm{L},k} =& 
  \big( \widehat{\bm{V}}_{\bm{\mathcal{H}}_k}^{\rm H} \bm{W}_k \widehat{\bm{V}}_{\bm{\mathcal{H}}_k} \big)^{-1}
  \widehat{\bm{V}}_{\bm{\mathcal{H}}_k}^\Hm \bm{W}_k \bm{D}_k \bm{F}_{\mathrm{AL},k}
  \widehat{\bm{V}}_{\mathrm{L},k} . \label{eq100}
\end{align}
 Given the optimal $\bm{\Sigma}_{\mathrm{L},k}$, the task is to find the optimal unitary
 matrix $\bm{V}_{\mathrm{L},k}$. For the optimization
\begin{align}\label{eq101}
\begin{array}{cl}
 \min\limits_{\bm{Q}} & \Vert \bm{B}\bm{Q} - \bm{A} \Vert_F^2 , \\
 \text{s.t.} & \bm{Q} \in \mathcal{U} ,
\end{array}
\end{align}
 the optimal solution is $\bm{Q}\! =\! \bm{U}\bm{V}^{\rm H}$
 \cite{inequalityMajorBook}, in which the unitary matrices $\bm{U}$ and $\bm{V}$ are
 given by the SVD $\bm{B}^{\rm H}\bm{A}\! =\! \bm{U}\bm{\Sigma}\bm{V}^{\rm H}$.
 Thus, for the optimization problem (\ref{prob-analog-org}), the optimal
 $\bm{V}_{\mathrm{L},k}$ is given by
\begin{align}\label{eq-unitary-matrix} 
 \bm{V}_{\mathrm{L},k} =& \bm{U}_{\bm{V}}^{\rm H}\bm{V}_{\bm{V}} ,
\end{align}
 where the unitary matrices $\bm{U}_{\bm{V}}$ and $\bm{V}_{\bm{V}}$ are defined
 based on the following SVD
\begin{align}\label{eq103}
 \big(\bm{V}_{\bm{\mathcal{H}}_k} \bm{\Sigma}_{\mathrm{L},k}\big)^{\rm H} \bm{W}_k
  \bm{D}_k \bm{F}_{\mathrm{AL},k} =& \bm{U}_{\bm{V}} \bm{\Sigma}_{\bm{V}} \bm{V}_{\bm{V}}^{\rm H} .
\end{align}

 Our analog beamformer design based on the weighted Frobenius norm minimization is very
 general. The different weight matrices can be utilized to realize different performance
 trade-offs. However, due to this weight matrix in the objective function, in most case,
 it is challenging to compute the analog beamformer in a closed-form. To overcome this
 difficulty, the analog beamformer optimization problem may be further transferred into the
 following optimization
\begin{align}\label{eq104}
 \min\limits_{\bm{F}_{\mathrm{AL},k}}  \ \ & \Big\lVert \bm{D}_k^{-1} \bm{V}_{\bm{\mathcal{H}}_k}
  \bm{\Sigma}_{\mathrm{L},k} \bm{V}_{\mathrm{L},k}^{\rm H} - \bm{F}_{\mathrm{AL},k} \Big\rVert_F^2 ,
\end{align}
 whose optimal solution can be directly computed by angle projection. In general,
 the weight matrix $\bm{W}_k$ is not an identity matrix, and in this case, the
 optimal solution of the analog beamformer is much more complicated. But it is
 also interesting to note that the solution of (\ref{eq104}) offers an `upper bound'
 to the general optimal analog transmit precoder as shown in \cite{matrixMonoOpt2018}.

\begin{algorithm}[!t]
	\caption{Analog Beamformer Optimization Algorithm}
	\label{ALG1}
	\begin{algorithmic}[1]
		\REQUIRE{Left singular matrix $\bm{V}_{\bm{\mathcal{H}}_k}$, weight matrix $\bm{W}_{k}$,
			invertible matrix $\bm{D}_k$, iteration threshold $\varepsilon$}
		\STATE Set initial objective function of (\ref{prob-analog-org}) to $\Delta =
		\varepsilon + 1> \varepsilon$
		\STATE Set initial $\bm{F}_{{\rm AL},k}\! =\! {\cal P}_{\cal F}\big(\bm{D}_k^{-1}
		\bm{V}_{\bm{\mathcal{H}}_k}\big)$, then calculate initial $\bm{V}_{{\rm L},k}$
		from SVD (\ref{eq93})
		\WHILE{$\Delta > \varepsilon$}
		\STATE{Update matrix $\widetilde{\bm{\Sigma}}_{\mathrm{L},k}$ and
			$\widehat{\bm{\Sigma}}_{\mathrm{L},k}$ using (\ref{eq-diag-left}) and (\ref{eq100})}
		\STATE{Compute unitary matrix $\bm{V}_{\mathrm{L},k}$ using (\ref{eq-unitary-matrix})}
		\STATE{Calculate analog beamformer matrix $\bm{F}_{\mathrm{AL},k}$ by solving
			(\ref{eq104}) using phase projection}
		\STATE{Compute $\Delta$ with new $\widetilde{\bm{\Sigma}}_{\mathrm{L},k}$,
			$\widehat{\bm{\Sigma}}_{\mathrm{L},k}$, $\bm{V}_{\mathrm{L},k}$ and $\bm{F}_{\mathrm{AL},k}$}
		\ENDWHILE
		\RETURN{Optimal analog beamformer $\bm{F}_{\mathrm{AL},k}$}
	\end{algorithmic}
\end{algorithm}

 Given $\bm{W}_{k}$ and $\bm{D}_k$, our algorithm to design the optimal analog
 transmit precoder for multi-hop AF relaying systems is summarized in
 Algorithm~\ref{ALG1}, where the objective function of (\ref{prob-analog-org}) is
 $\Delta\! =\! \big\Vert \bm{W}_k^{\frac{1}{2}} \big( \bm{V}_{\bm{\mathcal{H}}_k}
 \bm{\Sigma}_{\mathrm{L},k} \bm{V}_{\mathrm{L},k}^{\rm H}\! -\! \bm{D}_k
 \bm{F}_{\mathrm{AL},k} \big) \big\Vert_F^2$ \cite{matrixMonoOpt2018}.
 It is worth noting that concerning the high computational complexity of matrix inversion and singular value decomposition, the complexity of step 4 and step 5 is given by $ \mathcal{O}\bigl( ( N_{\rm RF} - N )^3 \bigr) $ and $ \mathcal{O}( N_{\rm RF}^3 ) $, respectively. Since matrix $ \bm{D} $ in (\ref{eq104}) does not change in iterations, it can be calculated off-line. Thus, for each iteration, the computational complexity of Algorithm 1 is $ \mathcal{O}( N_{\rm RF}^3 ) $. On the other hand, the calculation of analog precoders requires the knowledge of channel singular matrices. Therefore, the overall computational complexity of the proposed analog precoder design is $ \mathcal{O}( N_{r,k+1}^2 N_{t,k} + N_{t,k}^3 ) $ \cite{matrixComputationGolub2012}.

\subsubsection{Analog Receive Combiner Design}

 Next we look into the receiver design for hybrid transceivers. Based on
 Conclusions~\ref{C4}, \ref{C8} and \ref{C11}, the auxiliary variables
 $\bm{\Pi}_{{\rm R},k}$ for relays and $\bm{\Pi}_{\rm G}$ for destination can be
 unified into a single form. To be more specifically, according to the definition
 (\ref{eq53}), the receiving auxiliary matrix $\bm{\Pi}_{{\rm R},k}$ of
 (\ref{eq60}) for a relay node is
\begin{align}\label{eq105}
  \bm{\Pi}_{{\rm R},k} =& \big( {\bm{F}}_{{\rm AR},k} \bm{R}_{\bm{x}_{k-1}}
  {\bm{F}}_{{\rm AR},k}^{\rm H}\big)^{-\frac{1}{2}} {\bm{F}}_{{\rm AR},k} \bm{R}_{\bm{x}_{k-1}}^{\frac{1}{2}}.
\end{align}
 The auxiliary matrix of destination $\bm{\Pi}_{\rm G}$ can be obtained by
 simply substituting ${\bm{F}}_{{\rm AR},k}\bm{R}_{\bm{x}_{k-1}}^{1/2}$ with 
 $\bm{G}_{\rm A}$ in (\ref{eq105}). Thus, we only need to discuss the design of
 analog receive combiner for a relay node. According to Section~\ref{S5}, the
 main task here is to optimize the analog receive combiner ${\bm{F}}_{{\rm AR},k}$
 so that the right singular matrix of the auxiliary variables $\bm{\Pi}_{{\rm R},k}$
 can match the right singular matrix of the preceding-hop channel.
 
 Interestingly, it can be seen that by implementing the conjugate transpose
 operation on the both sides of (\ref{eq105}), the analog combiner design
 problem can be transformed into a similar form to the analog precoder design
 problem which has already been solved. Thus, by defining the SVD of the analog
 receive combiner
\begin{align}\label{eq106}
 \bm{F}_{{\rm AR},k}\bm{R}_{\bm{x}_{k-1}}^{1/2} = \bm{U}_{R,k} \bm{\Sigma}_{\mathrm{R},k}
  \bm{V}_{\mathrm{R},k}^{\rm H} ,
\end{align}
 and based on the previous work \cite{matrixMonoOpt2018}, the analog receive combiner
 design problem can be formulated as
\begin{align}\label{prob-analog-right} 
\begin{array}{cl}
 \min\limits_{\bm{\Sigma}_{\mathrm{R},k},\bm{U}_{\mathrm{R},k},\bm{F}_{\mathrm{AR},k}} & 
  \Big\lVert \bm{W}_k^{\frac{1}{2}} \big(\bm{U}_{\bm{\mathcal{H}}_k} \bm{\Sigma}_{\mathrm{R},k}
  \bm{U}_{\mathrm{R},k}^{\rm H}\! -\! \bm{R}_{\bm{x}_{k-1}}^{1/2} {\bm{F}}_{{\rm AR},k}^{\rm H}
  \big) \Big\rVert_F^2 , \\
 \text{s.t.} & \bm{U}_{\mathrm{R},k}\in \mathcal{U}, {\bm{F}}_{{\rm AR},k}\in \mathcal{F}_{\mathrm{PR},k},
  \ \bm{\Sigma}_{\mathrm{R},k} = \mathrm{diag} \big\{ \sigma_{\mathrm{R},1}, \cdots,
  \sigma_{\mathrm{R},K}\big\} . 
\end{array}
\end{align}

 Following the above discussions, it is clear that the analog receive combiner design
 for multi-hop AF relaying systems can also be completed using Algorithm~\ref{ALG1}.

\begin{algorithm}[!t]
\caption{Pseudo Algorithm for Hybrid Design for Multi-Hop Communications}
\label{ALG2}
\begin{algorithmic}[1]
 \REQUIRE{Channel matrices $\{\bm{H}_k\} $, maximum transmit powers of nodes $\{P_k\}$,
  standard deviations of noises at nodes $\{\sigma_{\bm{n}_k}\}$, positive semidefinite
  matrices $\{\bm{\Psi}_k\}$, maximum repeat counter $R$}
 \STATE Set repeat counter $r = 0$
 \STATE Calculate analog precoder $\bm{F}_{{\rm AL},1}$ at source (1st hop) based on
  \textbf{Conclusion~2}
 \STATE Calculate digital precoder $\bm{F}_{{\rm D},1}$ at source based on
  \textbf{Conclusion~3}
 \FOR{$k$th node ($1 < k < K$)}
   \STATE Compute analog combiner $\bm{F}_{{\rm AR},k}$ based on \textbf{Conclusion~4}
   \STATE Calculate analog precoder $\bm{F}_{{\rm AL},k}$ based on \textbf{Conclusion~6}
   \STATE Derive digital beamformer $\bm{F}_{{\rm D},k}$ according to \textbf{Conclusion~7}
 \ENDFOR
 \STATE Calculate analog combiner $\bm{F}_{{\rm AR},K}$ for $K$th node based on
  \textbf{Conclusion~8}
 \STATE Calculate analog precoder $\bm{F}_{{\rm AL},K}$ for $K$th node based on
  \textbf{Conclusion~10}
 \STATE Calculate analog combiner $\bm{G}_{\rm A}$ for destination based on
  \textbf{Conclusion~11}
 \STATE Compute digital beamformer $\bm{F}_{{\rm D},K}$ for $K$th node according
  to \textbf{Conclusion~12}
 \FOR{$k$th node ($1 < k \le K$) and repeat counter $r < R $}
   \STATE Update analog combiner $\bm{F}_{{\rm AR},k}$ based on latest $\bm{R}_{\bm{x}_{k-1}}$
   \STATE Update digital beamformer $\bm{F}_{{\rm D},k}$ based on latest $\bm{F}_{{\rm AR},k}$ 
   \STATE Update correlation matrix of received signal $\bm{R}_{\bm{x}_{k-1}}$ based on
    latest $\bm{F}_{{\rm D},k}$ and $\bm{F}_{{\rm AR},k}$.
 \ENDFOR
 \STATE Derive digital combiner $\bm{G}_{\rm D}$ for destination based on WMMSE criterion
 \RETURN{Analog and digital beamformer for each node in multi-hop communication}
\end{algorithmic}
\end{algorithm}

\subsubsection{Proposed Hybrid Transceiver Design}
 From the preceding discussions, it can be found that the multi-hop hybrid transceiver design is fundamentally different from the previous studies on point-to-point hybrid beamforming, e.g., \cite{WeiYuHybJSTSP2016,MYHongJSTSP2017Hybrid,JZhangJSTSP2016Alternating}, since there are basic differences on the system structures, resulting in, distinctive mathematical expressions between the multi-hop hybrid communications and the point-to-point hybrid communications. Correspondingly, the multi-hop hybrid communication system design faces different challenges to be addressed.
 Based on the conclusions given in Section~\ref{S5} and the above proposed analog
 design algorithm, the analog beamformer design problem for multi-hop communications
 can be solved directly. The pseudo algorithm for the designing procedure of our
 proposed hybrid transceiver optimization in the multi-hop communication system is summarized
 in Algorithm~\ref{ALG2}. Observe from Algorithm~\ref{ALG2} that because of the
 requirement of $\bm{R}_{\bm{x}_{k-1}}$ in the calculation of the analog combiner
 at the $k$th node as well as the unit-modulus constraints on the analog beamformers
 $\bm{F}_{{\rm AL},k}$ and $\bm{F}_{{\rm AR},k}$, an iterative process is necessary.
 Specifically, the analog combiner $\bm{F}_{{\rm AR},k} $, the digital beamformer
 $\bm{F}_{{\rm D},k}$ and the correlation matrix of the received signal
 $\bm{R}_{\bm{x}_{k-1}}$ at the relay are computed sequentially and repeatedly.
 It can be shown that for digital beamformer design the computational
 complexity is given by $ \mathcal{O}( N_{r,k+1}^2 N_{RF} ) $. Based on the
 complexity analysis for Algorithm 1, the complexity of Algorithm 2 can be directly
 written as $ \mathcal{O}( KRN_{r,k+1}^2 N_{t,k} + KRN_{t,k}^3 ) $. In particular,
 for a small number of hops, $ K $, and repeat counter, $ R $, the computational complexity
 of Algorithm 2 can also be formulated as $ \mathcal{O}( N_{r,k+1}^2 N_{t,k} + N_{t,k}^3 ) $.

\subsection{Unit-Modulus Alignment Design}\label{S6.2}

 The previous subsection has discussed an effective algorithm for the hybrid
 transceiver design, which considerably improves the achievable system's
 performance. This performance enhancement is of course achieved by sacrificing
 the computational complexity, see the loop of lines 13 to 17 in
 Algorithm~\ref{ALG2}. In practice, low complexity is also a major criterion
 measuring transceiver designs. Based on the results of Subsection~\ref{S6.1},
 a low complex hybrid transceiver design is proposed here to handle the
 unit-modulus beamformer design in multi-hop communications. 
 
 In particular, the analog beamformer design problem with the weight matrix chosen
 to be the identity matrix is considered, i.e., $\bm{W}_k\! =\! \bm{I} $. Again, we
 only need to discuss the analog transmit precoder design, since the analog receive
 combiner design problem can be obtained by transforming it into an analog precoder
 design. From Section~\ref{S5} and (\ref{eq-uniform}), it can be seen that the main
 task of the analog precoder design is to ensure the left singular matrix of
 $\bm{\Pi}_{{\rm L},k}$ corresponding to the eigenvectors of the channel
 $\bm{V}_{\bm{\mathcal{H}}_k}$. Noting the SVD of $\bm{D}_{k}\bm{F}_{\mathrm{AL},k}$
 given in (\ref{eq93}), the associated optimization problem can be formulated as
\begin{align}\label{eq108}
\begin{array}{cl}
 \min\limits_{\bm{F}_{\mathrm{AL},k}}\!\! &\!\!
  \Big\lVert \big( \bm{U}_{\mathrm{L},k} \bm{\Sigma}_{\mathrm{L},k}
  \bm{V}_{\mathrm{L},k}^{\rm H}\! -\! \bm{D}_k \bm{F}_{\mathrm{AL},k} \big) \Big\rVert_F^2 , \\
 \text{s.t.}\!\! &\!\! \bm{F}_{\mathrm{AL},k} \in \mathcal{F}_{\mathrm{PL},k} .
\end{array}\!
\end{align}
 However, as $\bm{F}_{\mathrm{AL},k}$ is tangled with positive semidefinite matrix
 $\bm{D}_k$, the problem (\ref{eq108}) is difficult to handle. Therefore, it is
 further transformed into its upper bound problem \cite{matrixMonoOpt2018}, which is 
\begin{align}\label{eq109}
\begin{array}{cl}
 \min\limits_{\bm{F}_{\mathrm{AL},k}}\!\! &\!\!
  \Big\lVert \big( \bm{D}_k^{-1} \bm{U}_{\mathrm{L},k} \bm{\Sigma}_{\mathrm{L},k}
  \bm{V}_{\mathrm{L},k}^{\rm H}\! -\! \bm{F}_{\mathrm{AL},k} \big) \Big\rVert_F^2 , \\
 \text{s.t.}\!\! &\!\! \bm{F}_{\mathrm{AL},k} \in
  \mathcal{F}_{\mathrm{PL},k} .
\end{array}\!
\end{align}
 Instinctively, the most effective way to obtain a unit-modulus matrix is to get the
 matrix's phase projection. Thus, the unit-modulus analog beamformer $\bm{F}_{\mathrm{AL},k}$
 is given by
\begin{align}\label{eq110}
 \bm{F}_{\mathrm{AL},k} = \mathcal{P}\left( \bm{D}_k^{-1} \bm{V}_{\bm{\mathcal{H}}_k} \right).
\end{align}

 Note that the analog receive combiner $\bm{F}_{{\rm AR},k}$ is still tangled with
 the correlation matrix of received signal $\bm{R}_{\bm{x}_{k-1}}$. Here, it is
 further assumed that $\{\bm{R}_{\bm{x}_{k-1}}\}$ are all identity matrices to
 simplify the design. With this assumption, the analog combiner $\bm{F}_{{\rm AR},k}$
 can be found in a similar way to (\ref{eq110}). The digital beamformer can be
 designed based on the results of Section~\ref{S5}.

 It is clear that there is no iteration involved in the proposed unit-modulus
 alignment design. Thus, compared with the algorithm of Subsection~\ref{S6.1},
 this algorithm has much lower computational complexity. This benefit is achieved
 by sacrificing the achievable system's performance.
 
\section{Numeral Results and Discussions}\label{S7}
 
 In order to assess the performance of the proposed solutions, several numerical
 results are presented. Without loss of generality, we investigate a three-hop AF
 MIMO relaying network. Unless otherwise stated, the source and destination are
 equipped with 32 antennas and 16 antennas, respectively, while there are 4 RF
 chains  involved in both the source and destination. The two relay nodes are both
 equipped with 32 antennas and 4 RF chains. From the source node, $N=4$ data streams
 are transmitted. It is worth noting that our derivation does not rely on a
 particular channel model. To demonstrate this, in the simulation both the millimeter
 wave (mmWave) channel and RF Rayleigh channel are considered. In the simulations,
 without loss of generality, the noise power is the same at every node, and the
 system's SNR is defined as the radio of the transmit signal power at source over
 the noise power at destination, i.e., $\text{SNR}=P_{\rm Tx}/\sigma_n^2$. In addition,
 the weight matrix in (\ref{prob-analog-org}) is set to be an identity matrix.

 Four hybrid transceiver designs with unit-modulus constraints are compared, and
 they are our proposed robust hybrid transceiver optimization design of
 Subsection~\ref{S6.1} (denoted by Proposed Alg.), our low-complexity unit-modulus
 alignment based design of Subsection~\ref{S6.2} (denoted as UMA Alg.), and two
 orthogonal matching pursuit (OMP) based designs \cite{HeathTWC2014,hybridRelay2014}.
 The OMP algorithm, originated from \cite{HeathTWC2014}, is widely used in
 point-to-point or one-hop hybrid transceiver designs, and it is extended to two-hop
 relay systems in \cite{hybridRelay2014}. To the best of our knowledge, the OMP
 algorithm applied to multi-hop ($K > 2$) scenarios has not been discussed in the
 existing literature. Based on the optimal structures presented in this paper and
 the previous discussions on the OMP algorithm given in \cite{HeathTWC2014,hybridRelay2014},
 we extend this algorithm to multi-hop scenarios. More specifically, we implement
 two OMP based algorithms in our simulations. The first one is referred to as the
 full digital based OMP algorithm, denoted as FD-OMP Alg., which is designed based
 on the optimal full digital solution as given in \cite{XingTSP2013}. The second one
 is known as the SVD based OMP algorithm, denoted as SVD-OMP Alg., which is designed
 based on singular matrices of channels. Specifically, define the SVDs of the
 channel matrices as
\begin{align}\label{eq111}
 \bm{H}_k =& \bm{U}_k \bm{\Sigma_k} \bm{V}_k^\Hm , ~ 1\le k\le K .
\end{align}
 Then the input beamformer required by the SVD-OMP Alg. for the $k$th node
 is given by
\begin{align}\label{eq112}
 \bm{F}_{{\rm in},k}^{\rm OMP} = \bm{V}_k \bm{U}_{k-1}^\Hm , 1\le k\le K ,
\end{align}
 where $\bm{U}_0\! =\! \bm{I}$. Then the traditional OMP algorithm is involved to
 compute analog and digital beamformers \cite{HeathTWC2014,hybridRelay2014}.
 The codebook of the two OMP algorithms for a mmWave channel is given by the
 channel steering vectors, and the codebook of the OMP algorithms for a RF
 channel is the same as that given in \cite{matrixMonoOpt2018}.  Furthermore,
 the powerful full digital transceiver design \cite{XingTSP2013} (denoted as
 Full Digital) is used as the ultimate benchmark. Note that for the full
 digital design, the number of RF chains must match the number of antennas.

\begin{figure}[!ht]
	\vspace*{-10mm}
	\begin{center}
		\includegraphics[width=0.7\textwidth]{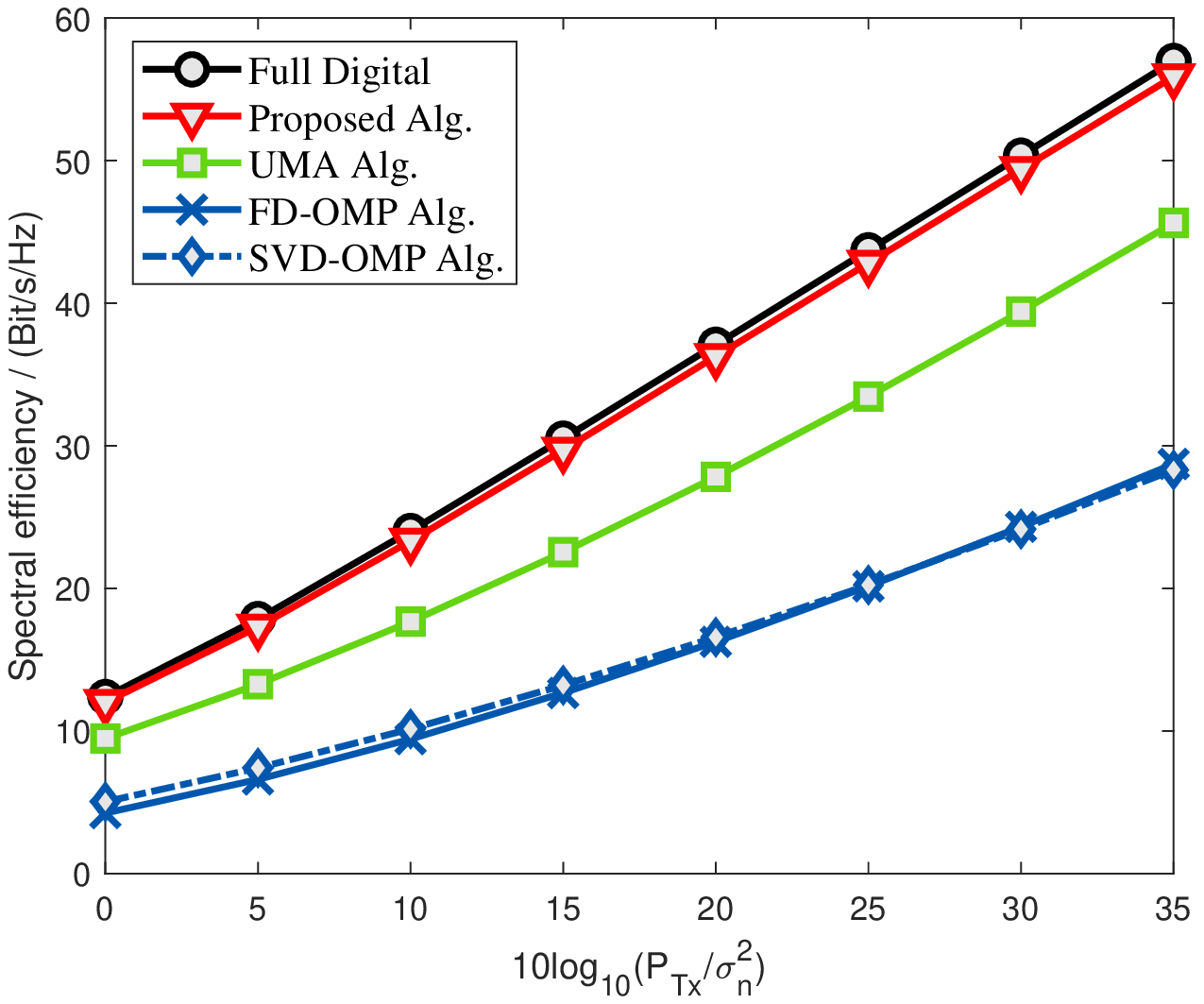}
	\end{center}
	\vspace*{-10mm}
	\caption{Comparison of spectral efficiency for the four linear hybrid transceiver
		designs and the full digital design based on capacity maximization. The mmWave
		channel with $N_{\rm path}=10$ paths is used in the simulation.}
	\label{FIG1}
	\vspace*{-6mm}
\end{figure}
 
 Initially, we assume that there is no channel estimation error for hybrid
 transceiver designs, and the results obtained are presented in Subsections~\ref{S7.1}
 and \ref{S7.2}. However, we also consider the case where the channel estimation
 error is not negligible in Subsection~\ref{S7.3}.

\subsection{MmWave Channel Case}\label{S7.1}

\begin{figure}[!b]
	\vspace*{-10mm}
	\centering
	\includegraphics[width=0.7\textwidth]{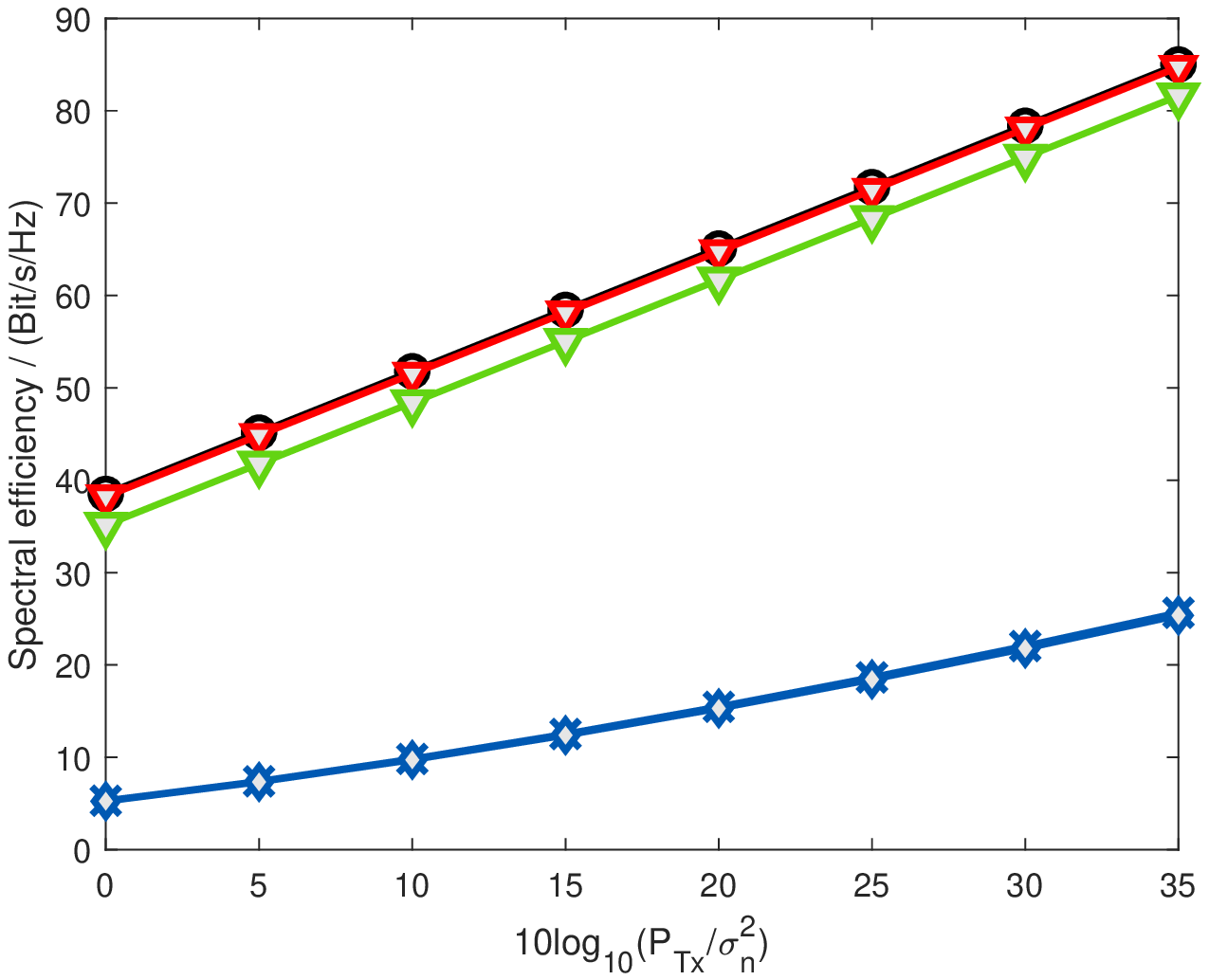}
	\vspace*{-8mm}
	\caption{Comparison of spectral efficiency for the four linear hybrid transceiver
		designs and the full digital design based on capacity maximization. The mmWave
		channel with $N_{\rm path}=10$ paths is used in the simulation.}
	\label{FIG2}
	\vspace*{-6mm}
\end{figure}

 Fig.~\ref{FIG1} compares the spectral efficiency performance of the four designs
 under the mmWave channel environment. Observe from Fig.~\ref{FIG1} that the
 performance of our proposed robust hybrid transceiver design is very close to
 the optimal performance of the full digital design, which is significantly better
 than the other four hybrid transceiver designs. It is worth noting that the
 performance of the both OMP Algorithms are equally poor for this three-hop AF
 relay MIMO system with $N_{\rm RF}\! =\! N$. This is contrast to the 
 traditional point-to-point MIMO systems, where the original OMP algorithm performs
 well \cite{HeathTWC2014,hybridRelay2014}. The results of Fig.~\ref{FIG1} therefore
 show that the hybrid transceiver design based on the OMP Algorithm is not suitable
 for complicated multi-hop communication systems. As expected, the simulation results
 indicate that our proposed lower-complexity UMA Alg. suffers observable performance
 loss compared to our Proposed Alg. but crucially, it significantly outperforms the
 two OMP Algorithms. This indicates that if the  low complexity is a critical
 performance measure, our UMA Alg. offers a suitable design choice.

\begin{figure}[!b]
	\vspace*{-8mm}
	\begin{center}
		\includegraphics[width=0.7\textwidth]{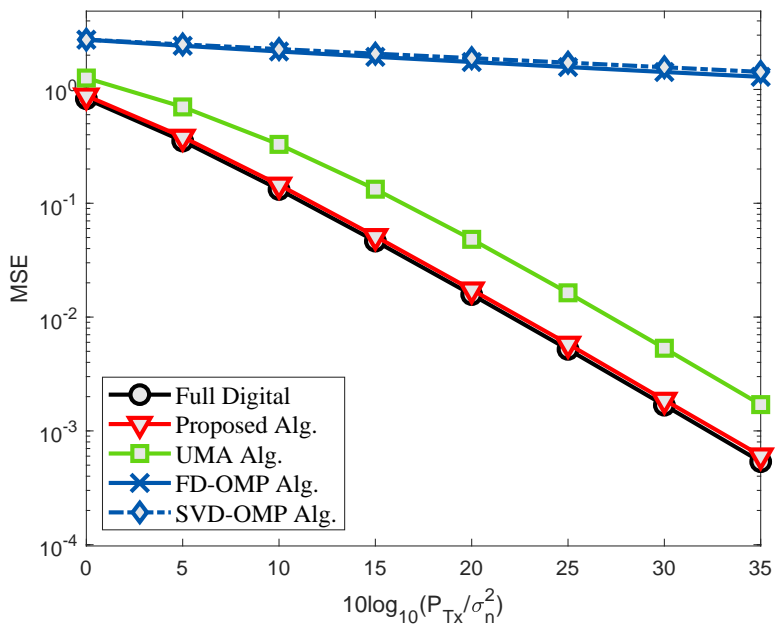}
	\end{center}
	\vspace*{-10mm}
	\caption{Comparison of transmitted signal MSE for the four linear hybrid transceiver
		designs and the full digital design based on MSE minimization. The mmWave channel
		with $N_{\rm path}=10$ paths is used in the simulation.}
	\label{FIG3}
	\vspace*{-6mm}
\end{figure}

 Further, in Fig.~\ref{FIG2}, the comparison of the spectral efficiency of the five
 designs in the massive MIMO communications is presented. MmWave channel is utilized
 in the simulation. The source, two relay nodes, and the destination are all equipped
 with $ 256 $ antennas. From Fig.~\ref{FIG2}, it can be seen that the performance of
 the proposed algorithm outperforms the other four hybrid transceiver designs,
 which is nearly the same with the full digital transceiver. It is also shown in
 Fig.~\ref{FIG2} that the sum rate of the UMA Alg. is quite close to that of the full
 digital design. This is because the extra antennas provide more spatial diversity
 and multiplexing gain for the hybrid transceivers. In addition, the spectral
 efficiency of the two OMP algorithms improves as well, but still falls largely
 behind of the proposed hybrid design. Therefore, the result indicates that the
 proposed algorithm retains its superiority in the channel scenario with 256 antennas,
 and it also shows the capability of the proposed hybrid design in massive MIMO channels.
 
 Fig.~\ref{FIG3} compares the MSE minimization performance of the five designs
 under the mmWave channel senario. The results obtained again demonstrate that
 the achievable performance of our robust hybrid transceiver design is very
 close to the powerful full digital design, namely, it is near optimal, while
 imposing substantially lower hardware costs in comparison with the optimal full
 digital design. The results of Figs.~\ref{FIG3} also show that hybrid
 transceiver design based on the two OMP Algorithms are similarly very poor, and
 this indicates that the OMP based design is not suitable for multi-hop AF MIMO
 relaying networks under the MSE minimization criterion too. Observe that our UMA
 Alg. is capable of achieving considerably lower computational complexity by
 trading off some achievable MSE performance, compared with our near optimal
 robust hybrid transceiver design. In particular, it outperforms the OMP based
 design considerably.

\begin{figure}[!t]
\vspace*{-6mm}
\begin{center}
\includegraphics[width=0.7\textwidth]{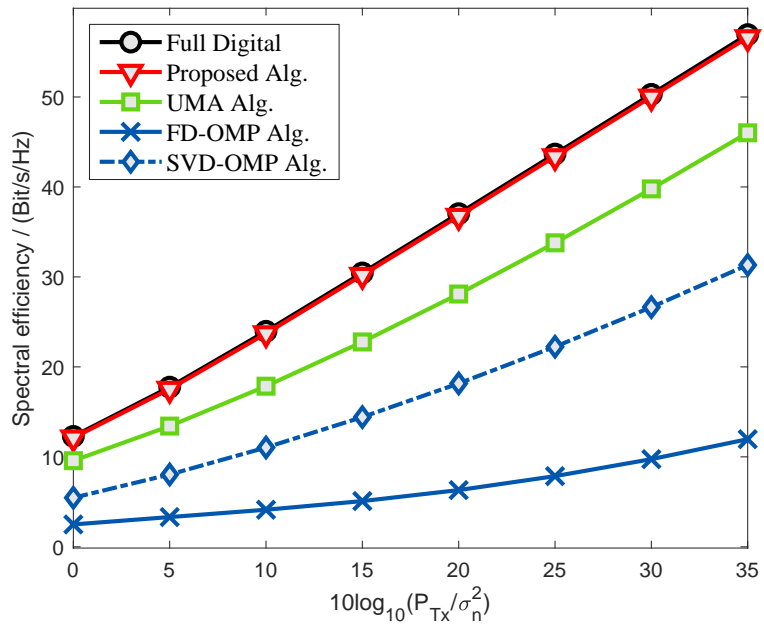}
\end{center}
\vspace*{-10mm}
\caption{Comparison of spectral efficiency for the four linear hybrid transceiver
 designs and the full digital design based on capacity maximization.  The mmWave
 channel with $N_{\rm path}=10$ paths is used in the simulation, and the number of
 RF-chains in each node is set to $N_{\rm RF}=6$.}
\label{FIG4}
\vspace*{-6mm}
\end{figure}

 In addition, the communication scenario with extra RF-chains is tested in Fig.~\ref{FIG4}.
 Specifically, in the simulation, the number of data streams remains $N\! =\! 4$, and we
 set the number of RF-chains to $N_{\rm RF}\! =\! 6$, which means there are 2 extra
 RF-chains that can be used to enhance the system performance. By comparing Fig.~\ref{FIG4}
 with Fig.~\ref{FIG1}, it can be seen that with more RF-chains than data streams, the
 hybrid transceiver system performance can indeed be improved. In particular, from
 Fig.~\ref{FIG4}, it is clear that the performance of the Proposed Alg. is almost identical
 to that of the optimal full digital design, and the performance of the UMA Alg. is also
 improved slightly. Furthermore, the performance of the SVD-OMP Alg. is also improved,
 but it remains significantly worst than our low-complexity UMA Alg. design. However,
 for the FD-OMP Alg., increasing $N_{\rm RF}$ to more  than $N$ actually degrades the
 system performance considerably. The reason is  as follows. The need of $\bm{R}_{\bm{x}_k}$
 in the FD-OMP Alg., required by the input full digital solution, naturally leads to
 mismatch between the optimal full digital transceiver and the FD-OMP Alg. based transceiver.
 The extra RF chains will magnify this mismatch, and results in a worse performance.

\begin{figure}[!t]
\vspace*{-6mm}
\begin{center}
\includegraphics[width=0.7\textwidth]{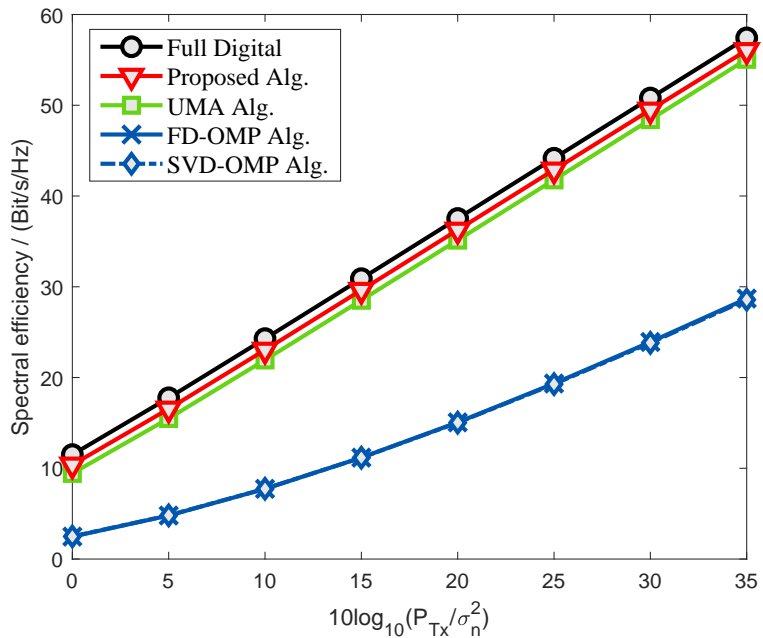}
\end{center}
\vspace*{-10mm}
\caption{Comparison of spectral efficiency for the four linear hybrid transceiver
 designs and the full digital design based on capacity maximization. The RF
 Rayleigh channel is used in the simulation.}
\label{FIG5}
\vspace*{-6mm}
\end{figure}

\subsection{RF Channel Case}\label{S7.2}

 The simulation results for the RF Rayleigh channel are depicted in Fig.~\ref{FIG5}.
 It can be seen that under the RF Rayleigh channel environment, our robust hybrid
 transceiver design only suffers from very slight performance degradation, in
 comparison to the optimal full digital design. Furthermore, our low-complexity 
 UMA Alg. now attains a performance close to that of the Proposed Alg., since
 the rich scatters in the RF channel environment provide much more tolerance to
 mismatch between the theoretical optimal analog beamformer and the actual analog
 beamformer.  It is worth pointing out again that the two OMP Algorithms have a
 design challenge for Rayleigh channels, owing to the lack of steering vector
 based codebooks. Therefore, similar to \cite{matrixMonoOpt2018}, we have to use
 the phase matrix of the Rayleigh channel as the OMP codebook.

\subsection{Robust Design with Channel Estimation Error}\label{S7.3}

\begin{figure}[!t]
	\vspace*{-6mm}
	\begin{center}
		\includegraphics[width=0.7\textwidth]{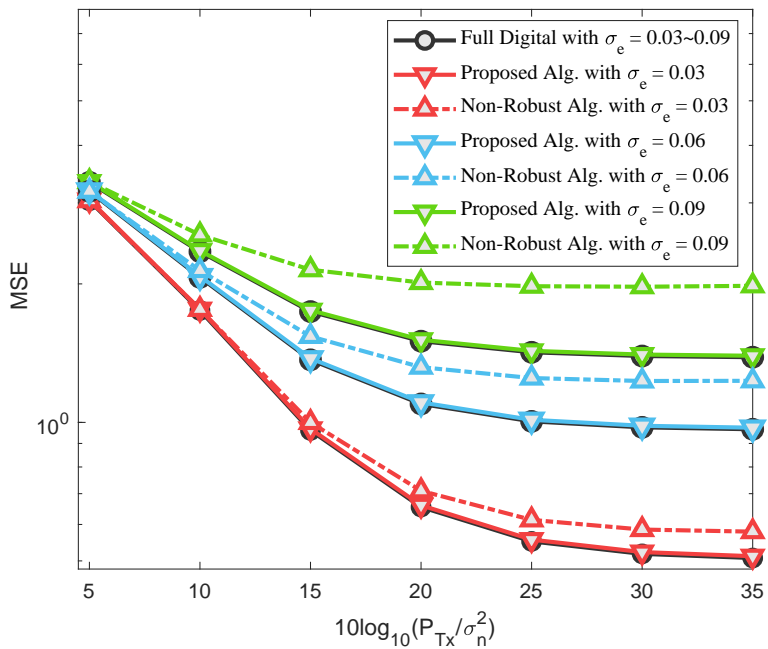}
	\end{center}
	\vspace*{-10mm}
	\caption{Comparison of transmitted signal MSE under different channel estimation
		errors for the proposed robust hybrid transceiver design, the full digital robust
		transceiver design, and non-robust hybrid transceiver design based on MSE
		maximization with $\alpha_e=0.6$. The RF Rayleigh channel is used in the simulation.}
	\label{FIG6}
	\vspace*{-7mm}
\end{figure}

 In the above simulation investigation, there exists no channel estimation error
 in hybrid transceiver designs. In practice, however, the channel estimation
 error always exists and cannot be neglected. Thus, the channel estimation error
 is considered. Specifically, the transmit correlation matrix of the channel error
 is defined based on the exponential model with the $i$th-row and $l$th-column
 element of the correlation matrix $\bm{\Psi}_k$ given by
\begin{align}\label{eq113}
 [\bm{\Psi}_{k}]_{i,l} =& \sigma_{e,k} \alpha_{e,k}^{\lvert i - l \rvert} .
\end{align}
 Without loss of generality, it is assumed that the correlation coefficients
 $\lbrace \sigma_{e,k} \rbrace$ and the variances $\lbrace \alpha_{e,k} \rbrace$
 of the transmit correlation matrix of channel error are the same for every
 channel, and they are denoted by $\sigma_e$ and $\alpha_e$, respectively. In 
 the simulation, $\alpha_e=0.6$ is adopted. We consider our robust hybrid
 transceiver design, Proposed Alg., under this imperfect channel condition. 
 Additionally, the counterpart of our Proposed Alg., which treats the estimated
 but inaccurate channel as the perfect one \cite{XingTSP201502}, is used for
 comparison, and it is denoted as Non-Robust Alg.

 Only the RF Rayleigh channel senario is considered. Fig.~\ref{FIG6} compares the
 performance of the proposed robust hybrid transceiver design, the full digital
 robust transceiver design \cite{XingTSP2013} and non-robust hybrid transceiver
 design \cite{XingTSP201502}, in terms of MSE, under different channel estimation
 errors. From Fig.~\ref{FIG6}, it can be seen that the proposed robust hybrid
 transceiver design is very close to that of the optimal full digital robust
 transceiver design, and  it achieves better performance than the non-robust
 hybrid transceiver design. Moreover, as the channel estimation error, i.e.,
 $\sigma_e$, increases, the performance gap between our proposed robust hybrid
 transceiver design and the  non-robust hybrid transceiver design becomes larger.

\section{Conclusions}\label{S8}

 In this paper, we have investigated robust hybrid transceiver optimization for
 multi-hop AF MIMO relaying networks, in which all nodes employ hybrid transceivers
 and multiple data streams are transmitted from source node simultaneously. A
 unified design framework has been proposed for both hybrid linear and nonlinear
 transceivers under generic objective functions, which also takes into account
 channel estimation error. Based on the proposed framework, it has been shown that
 the analog transceivers and digital transceivers can be decoupled without loss of
 optimality. Using matrix-monotonic optimization framework, the optimal structures
 of the analog and digital transceiver designs have been derived, which greatly
 simplify the hybrid transceiver optimizations. Based on the derived optimal structures,
 both analog precoders and combiners as well as digital forward matrices can be optimized
 separately and efficiently. Simulation results obtained have demonstrated that
 our proposed robust hybrid transceiver design only suffers from a very slight
 performance loss compared to the powerful full digital design. This confirms that
 our hybrid transceiver design attains near optimal performance, while imposing
 substantially lower hardware cost than the full digital design.

\bibliographystyle{IEEEtran}

\end{document}